\PassOptionsToPackage{pdfpagelabels=false}{hyperref}
\documentclass[useAMS,fleqn,usenatbib]{mnras}
\pdfoutput=1 
\usepackage{graphicx}
\usepackage[dvipsnames]{xcolor}
\usepackage{amsmath,amssymb,amstext}
\usepackage[T1]{fontenc}
\usepackage{ae,aecompl}
\usepackage[utf8]{inputenc}
\usepackage{newtxtext,newtxmath}
\usepackage[figure,figure*,table,table*]{hypcap}
\include{hyperlink-year-only-natbib-patch}

\newcommand*{\kms}{\ensuremath{\text{km}\,\text{s}^{-1}}}
\newcommand*{\msun}{\ensuremath{\text{M}_\odot}}
\newcommand*{\perh}{\ensuremath{h^{-1}}}

\newcommand*{\mailto}[1]{\href{mailto:#1}{#1}}
\newcommand*{\http}[1]{\href{http://#1}{#1}}

\newcommand*{\donearz}[1]{}
\newcommand*{\donecef}[1]{}
\newcommand*{\doneyao}[1]{}

\title[Predictably Missing Satellites]{Predictably Missing Satellites: Subhalo Abundances in Milky Way-like haloes}

\author[C.~E.~Fielder et al.]{%
Catherine~E.~Fielder\thanks{E-mail: \mailto{cef41@pitt.edu}},
Yao-Yuan~Mao\thanks{E-mail: \mailto{yymao.astro@gmail.com}, \mailto{yymao@pitt.edu}},
Jeffrey~A.~Newman,
Andrew~R.~Zentner,
\newauthor
and Timothy~C.~Licquia
\\
Department of Physics and Astronomy and the Pittsburgh Particle Physics, Astrophysics and Cosmology Center (PITT PACC),\\ 
University of Pittsburgh, Pittsburgh, PA 15260, USA\\
}

\pubyear{2018}
\begin{document}
	
\label{firstpage}
\pagerange{\pageref{firstpage}--\pageref{lastpage}}
\maketitle

\begin{abstract}
On small scales there have been a number of claims of discrepancies between the standard Cold Dark Matter (CDM) model and observations. The 'missing satellites problem' infamously describes the over-abundance of subhaloes from CDM simulations compared to the number of satellites observed in the Milky Way. A variety of solutions to this discrepancy have been proposed; however, the impact of the specific properties of the Milky Way halo relative to the typical halo of its mass have yet to be explored. Motivated by recent studies that identified ways in which the Milky Way is atypical, 
we investigate how the properties of dark matter haloes with mass comparable to our Galaxy's --- including concentration, spin, shape, and scale factor of the last major merger --- correlate with the subhalo abundance. Using zoom-in simulations of Milky Way-like haloes, we build two models of subhalo abundance as functions of host halo properties. From these models we conclude that the Milky Way most likely has fewer subhaloes than the average halo of the same mass. We expect up to $30\%$ fewer subhaloes with low maximum rotation velocities ($V_{\rm max}^{\rm sat} \sim 10$km\,s$^{-1}$) at the 68\% confidence level and up to 52\% fewer than average subhaloes with high rotation velocities ($V_{\rm max}^{\rm sat} \gtrsim 30$km\,s$^{-1}$, comparable to the Magellanic Clouds) than would be expected for a typical halo of the Milky Way's mass. Concentration is the most informative single parameter for predicting subhalo abundance.  Our results imply that models tuned to explain the missing satellites problem assuming typical subhalo abundances for our Galaxy may be over-correcting.
\end{abstract}


\section{Introduction}
\label{Section:intro} 

Simulations of structure formation based upon the $\Lambda$CDM (cold dark matter with a cosmological constant) cosmological model successfully describe a wide range of observations, particularly at large scales ($r \gtrsim 10$\,Mpc\,\perh). However, a number of observations on smaller scales ($r \lesssim 1$\,Mpc\,\perh) exhibit possible discrepancies with the standard model \citep[e.g.,][]{flores1994,moore1994,klypin1999,moore1999,b-k2011,delpopolo2017,bullock2017}. Many of these discrepancies have been observed in the Milky Way (MW), the galaxy we are able to study in the most detail. 

In this paper we focus most directly on an issue known as the 
`missing satellites problem' (MSP). The MSP is the apparent over-prediction of the abundance of satellite haloes of a particular velocity dispersion (or rotation speed) within a $\Lambda$CDM model relative to the number of galaxies of similar velocities that have actually been observed in the Milky Way \citep[e.g.,][]{klypin1999,moore1999,bullock2010}. We expect substructure to survive the hierarchical assembly of dark matter (DM) haloes, as the dense cores of merging haloes are not strongly affected by tidal interactions \citep{kauffmann1993,zentner2003,bullock2010} but the question of how much substructure there should be remains.

We describe objects in terms of maximum circular velocity, $V_{\rm max}$, which is a measure of the depth of the potential well of a dark matter halo. The critical range of maximum circular velocity we are interested in for studying the MSP is the range of $10-30$\,\kms, where other physical effects do not dominate (gas cooling on the low velocity end and photoionization feedback on the high velocity end). There has only been $\sim 26$ dwarf satellites observed of this size, compared to roughly $\sim 350$ predicted from CDM simulations \citep{tollerud2008,bullock2010} with $V_{\rm max} > 10$\,\kms within 400\,kpc of a Milky Way-size host. Of these $\sim 100$ should be massive enough to host observable galaxies \citep{bullock2010}.  The MSP is the difference between the number of predicted subhaloes of the size necessary to host galaxies with the observed stellar kinematics of the classical Milky Way dwarf satellites and the number of actual dwarf satellites. Within the $\Lambda$CDM framework, most such subhaloes cannot host dwarf satellites comparable in size to the classical MW dwarfs. Many previous works have tried to understand this discrepancy 
via some baryonic processes (e.g., supernova feedback) or through exotic physics (e.g., alternative models of dark 
matter).

The missing satellites problem is not the only example of potential small-scale discrepancies from $\Lambda$CDM expectations found in the Local Group. The simplest possible relationship between galaxies and dark matter would place the known satellites of the Milky Way into the largest subhaloes surrounding it \citep{bullock2017}. The 'too-big-to-fail’ (TBTF) problem was identified based upon $\Lambda$CDM N-body simulations \citep{springel2008,diemand2008} which showed that the most massive subhaloes were too dense to host the the brightest Milky Way satellites \citep{b-k2011,b-k2012}. Naively, these dense subhaloes should form stars more efficiently than their more diffuse counterparts, yet they do not appear to host the Local Group dwarfs \citep{tollerud2014,kirby2014}. 

A number of solutions have been proposed to alleviate or eliminate the MSP and TBTF problems. Some of the most well-studied proposals include:

\begin{enumerate}

\item \textbf{Survey incompleteness:} The missing objects could be too dim or too diffuse to have been detected yet. This question could partially be reconciled with future survey projects such as LSST, a hope that has gained traction after the numerous satellite galaxy discoveries from SDSS and DES \citep{blanton2017,drlica2015}, although most of these dwarfs are predominately much smaller than the regime that we are interested in ($V_{\rm max} \lesssim 10$ km\,s$^{-1}$). Recent work by \citet{kim2018} argues that there is no longer any MSP based upon our current understanding of the suppression of star formation in the satellites and the recent discovery of faint dwarfs after being completeness corrected to the Milky Way's virial radius. They predict that the number of satellites that inhabit the Milky Way is consistent with CDM predictions. However, there is still some issue with objects smaller than Segue I ($V_{\rm max} = 10^{+7.0}_{-1.6}$ km\,s$^{-1}$ from \citet{jiang2015}) in their model being over-predicted. In this case, the observed abundance of satellites can be explained, but the question of why subhaloes of similar potential well depth may host galaxies of very different luminosities would still be open. 

\item \textbf{Host halo properties:} \citet{wang2011} pointed out that the Milky Way's mass is uncertain by roughly a factor of two; if the MW mass is on the low side of this range, then both the missing satellites and TBTF problems are greatly alleviated.
In addition, several studies have shown that there is significant scatter in subhalo abundances among host haloes at fixed host halo mass \citep{purcell2012,jiang2014,vandenbosch2014}. Both \citet{purcell2012} and \citet{jiang2014} argued that the TBTF problem is only marginally significant given this large scatter, even if the mass of the Milky Way halo is on the high end of the observed range. 

\item \textbf{Baryonic effects:} Both MSP and TBTF are issues that arise when comparing observations with $N$-body simulations. However, baryonic physics can potentially change the observable satellite population substantially. Galaxies in small subhaloes may be dim because their host haloes did not form stars significantly prior to reionization, at which point their gas was stripped by the UV background \citep{bullock2000,benson2002,moore2006,bovill2009}. Milky Way subhaloes may have experienced a stronger suppression caused by the radiation from the MW itself. Photoionization is expected to affect haloes in the range of $V_{\rm max} \sim 30 \,\rm kms^{-1}$ \citep[e.g.,][]{efstathiou1992,bullock2000,benson2002,bovill2009,sawala2016,bullock2017}. 

Likewise, supernova feedback heats and blows out the interstellar medium, which can suppress star formation in low-mass dark matter haloes \citep{dekel1986,mori1999,scannapieco2008,koposov2009}. This process generally effects larger subhaloes of $V_{\rm max} \sim 100$ km\,s$^{-1}$. Winds and UV radiation from massive stars may also deposit energy into the gas within a subhalo, enhancing this effect \citep{ceverino2009}. 

Hydrodynamic simulations have backed the baryon solution up, showing that adding baryonic physics into dark matter-only models can substantially alleviate the missing satellites problem \citep[e.g.,][]{zolotov2012,sawala2016,zhu2016}. For example, dynamical friction resulting in angular momentum transfer from baryons to dark matter alters the dark matter density profile at the centres of low-mass galaxies \citep{brooks2013}, making them more cored than cuspy. These objects have less visible stars due to stronger effects of tidal stripping on cored density profiles \citep{delpopolo2009}. 

Tidal stripping by the host halo or the parent galaxy during accretion could have torn apart some fraction of the lower mass satellites \citep{diemand2007,bland2016,garrisonkimmel2017,nadler2019}. 

\item \textbf{Non-cold dark matter and exotic physics:} Dark matter properties which deviate from CDM could reduce the  number of subhaloes of appropriate potential well depth to host the observed satellite galaxies, thus alleviating or eliminating the MSP and the TBTF issue \citep[e.g.,][]{sommer2001,spergel2000,zentner2003,wang2014}. Additionally \citet{kamionkowski2000} and \citet{zentner2003} argue that non-standard inflationary scenarios could alter the initial conditions for structure formation, yielding fewer satellites at lower mass scales and alleviating  
small-scale issues such as the MSP. 
\end{enumerate}

Among these possible solutions, the study of the impact of host halo properties has mostly been limited to host halo mass. In $N$-body simulations, including the original simulation from which the haloes in this work were drawn, the abundance of resolved subhaloes is found to be directly proportional to the mass of the host halo over several orders of magnitude in mass \citep[e.g.,][]{gao2004,kravtsov2004,boylan-kolchin2010}. However, other halo properties, such as halo concentration and spin, correlate with the subhalo population as well; incorporating their effects will yield improved predictions of subhalo populations. This will enable more direct comparisons of the satellite abundances around the Milky Way to the subhalo populations of a halo \textit{whose properties match our Galaxy's.} 

In this paper, our first goal is to use high resolution $\Lambda$CDM zoom-in simulations for Milky Way-mass haloes (described in \autoref{sub:simulations}) to explore the nature of the host-to-host scatter in subhalo populations. Our study seeks, in part, to determine which host halo properties determine the distribution of subhalo abundance at fixed host halo mass. We then make predictions for substructure abundances in the Milky Way which incorporate observational constraints on the properties of its host dark matter halo, and explore the impact on the missing satellites problem. 


A variety of data has provided constraints on various properties of the dark matter halo which hosts the Milky Way other than its total mass. For example, kinematics of halo stars \citep[e.g.,][]{deason2012} and masers \citep{nesti2013} have constrained the Milky Way halo concentration by constraining its mass distribution, while the tidal stream of the Sagittarius dwarf \citep{law2010,veraciro2013} has constrained the halo's shape. We may grossly infer its angular momentum as well. For example, \citet{licquia2016} show that the Milky Way's stellar disk has a scale length roughly a factor of two lower than would be typical given its mass (or luminosity) and rotation speed, lying further from the luminosity-velocity-radius relation than roughly 90$\%$ of spirals. This small scale length would be expected to be related to its halo spin parameter \citep{mmw1998a}.
As a result, the atypical scale length of the Milky Way suggests that its dark matter halo may also be unusual; we wish to explore any impact this has on the abundance of subhaloes in our Galaxy.

The ``tailor-made'' prediction of Milky Way subhalo abundance which we obtain in this paper is relevant to our interpretation of the small-scale challenges to $\Lambda$CDM in many ways. Suppose, for instance, that we assume the most conservative scenario in which any small-scale issues can be completely resolved by invoking baryonic processes rather than by introducing any new physics. We will show in this paper that it is reasonable to expect that the Milky Way host halo includes roughly one-fourth fewer than average smaller subhaloes and as much as three-fifths fewer than average larger subhaloes than would be typical given its mass, based upon the correlations of satellite abundance with other Milky Way halo properties. In that case, the impact of baryonic physics on the MSP and TBTF problems must be significantly smaller than has been assumed in the past (as otherwise we have observed \textit{more} satellites around the Milky Way than would be expected in $\Lambda$CDM). 

This has important consequences for the tuning of the parametrized models of baryonic physics used in simulating galaxy evolution. For example, if we suppose that the MSP is resolved largely by baryonic feedback, then this feedback may need to be significantly less efficient than previously thought, since the Milky Way should be expected to have fewer subhaloes to begin with than a typical halo of its mass. Such alterations to baryonic physics models may have extensive impacts on simulations of dwarf galaxy formation and of galaxy evolution more generally. In addition, our study also provides a theoretical context for interpreting the observed satellite luminosity functions of Milky Way-like hosts that are outside the Local Volume \citep{geha2017}, with which we can test how impactful host halo properties are on the satellite populations by correlating them with proxies for halo properties such as disk scale length. The general strategy that we describe and advocate in this manuscript will become increasingly 
useful as more and more becomes known about the Milky Way galaxy and the halo 
of the Milky Way.

The structure of this paper is as follows.  First, in \autoref{Section:milkyway} we describe the basic host dark matter halo properties we focus on in this paper -- spin, shape, concentration, and merger history -- and describe what is known about each for the Milky Way. In \autoref{section:satellites}, we show that host halo properties are strongly correlated with the total abundance of subhaloes above a threshold in circular velocity. We then examine what model based upon the host halo properties provides the best predictions for satellite abundance for Milky Way-like hosts, and evaluate that model using the observed properties of the Milky Way host halo. In \autoref{Section:conclusion} we conclude that due to the somewhat unusual formation history of the Milky Way's host halo, we expect that it should have fewer subhaloes than typical for its mass, and discuss some implications and caveats. Additionally, in the first Appendix (\autoref{Section:Adiabatic Contraction}) we explain the details of obtaining a rough concentration estimate for the Milky Way when incorporating adiabatic contraction, and in the second Appendix (\autoref{Section:Numerical Techniques}) we discuss in detail the numerical and mathematical techniques utilised in this work and provide fitting functions for estimating subhalo abundance based upon host halo properties.

\section{Milky Way Halo Properties}
\label{Section:milkyway}

The following subsections detail the host halo properties we use in our analyses and include approximate estimates for the Milky Way. We stress that the values for the Milky Way are subject to both measurement and modelling errors. 
The most important take away is the approximate rank of the Milky Way in order to compare it to other dark matter haloes.

\subsection{Zoom-in Simulations}
\label{sub:simulations}

In the analyses presented here, we use a set of zoom-in cosmological simulations consisting of 45 Milky Way-mass haloes. These haloes were selected from a 125\,Mpc\,\perh parent simulation containing 1024$^{3}$ particles. The cosmological parameters for the simulations are $\Omega_{\rm M} = 0.286$, 
$\Omega_{\Lambda} = 1 - \Omega_{\rm M} = 0.714$, $h=0.7$, mass fluctuation amplitude $\sigma_{8} = 0.82$, and scalar spectral index $n_{\rm s} = 0.96$. All of the Milky Way-analog haloes selected for re-simulation fall within the mass range of 
$M_{\rm vir} =  10^{12.1 \pm 0.03}\ \mathrm{M}_{\odot}$. The mass of the highest-resolution particles in the zoom-in simulations is 
$m_{\rm p} = 3.0 \times 10^{5}\;\mathrm{M}_{\odot}\,\perh$. The softening length within the highest-resolution region is 170\,pc\,\perh comoving. The lower limit in $V_{\rm max}$ for convergence is approximately 10\,\kms. For more details on the simulation suite, refer to \citet{mao2015}. 

We use the {\tt ROCKSTAR} halo finder to identify haloes and subhaloes within each simulation. Halo masses and radii are defined as virial values using the virial overdensity threshold $\Delta_{\rm vir}$, which has a value of $\approx 340$ given the cosmological parameters from the previous paragraph. Subhaloes are haloes whose centres lie within the virial radius ($R_{\rm vir}$) of a halo that has a larger maximum circular velocity, $V_{\rm max} =$ max$[GM(<R)/R]^{1/2}$. We refer to any halo that is not a subhalo as a host halo. In each simulation, we select every subhalo that lies within 200 kpc of the host halo's centre and has $V_{\rm max} > 10 \,\text{km}\,\text{s}^{-1}$. Generally in this paper, we use $V_{\rm max}$ as a measure of the potential well depth in a subhalo. In the following, we will often quote subhalo circular velocities in units of the maximum circular velocity of the host halo, namely $V^{\rm sat}_{\rm max}/V^{\rm host}_{\rm max}$, because subhalo demographics are approximately self-similar when scaled in this way. Our resolution limit corresponds to a limit on this ratio of $ V^{\rm sat}_{\rm max}/V^{\rm host}_{\rm max} > 0.065$.

Halo properties (e.g., concentration, spin, and so on; see below) are all computed as described in the {\tt ROCKSTAR} documentation \citet{behroozi2013}.

\subsection{Concentration}
\label{sub:concentration}

On average, CDM haloes can be described by a universal density profile which is approximated well by the Navarro, Frenk, \& White (NFW) profile of \citet{nfw1996}. The NFW profile can be written as 
\begin{equation}
\rho(r) = \frac{\rho_{\rm char}}{\frac{r_{\rm vir}}{r_{\rm s}}(1+\frac{r_{\rm vir}}{r_{\rm s}})^{2}},
\end{equation}
where $\rho_{\rm char}$ is the characteristic overdensity, $r_{\rm vir}$ is the virial radius, and $r_{\rm s}$ is the scale radius.
The scale radius of a halo is most often expressed through the concentration 
parameter $c_{\mathrm{NFW}}$, which is the ratio of the halo virial radius to the halo scale radius, 
\begin{equation}
c_{\rm NFW } = \frac{r_{\rm vir}}{r_{\rm s}}.
\label{eq:NFW}
\end{equation}
The concentration parameter characterises the degree to which the mass of the halo is concentrated toward the halo centre. Concentration is known to be a slowly-declining function of halo mass \citep{nfw1997, bullock2001}; it has previously been found to correlate with subhalo abundance \citep{mao2015}. 

We can compare the concentration of simulated haloes in the mass range of the Milky Way to the approximate constraints of the Milky Way halo concentration, along with the haloes of galaxies of the same Hubble type (SBb/c in the case of the Milky Way; \citealt{bullock2001,nfw1997}). $\Lambda$CDM models of $10^{12}\rm M_{\odot}$ haloes have placed $c_{\rm NFW}$ in the range of 11--21. However, the constraint on ``normal'' Sb galaxies according to \citet{klypin2002} is expected to be $\sim$ 10--17, based on statistical ensembles of haloes. It is challenging to pin down an exact estimate of concentration for Milky Way-sized haloes in dark matter-only simulation, as baryons are expected to cause haloes to adiabatically contract and drive up their concentrations \citep[][]{eggen1962,blumenthal1986,mmw1998a,cintio2014}, while feedback and/or mergers can reduce halo concentrations. Concentration measurements of the range 18--24 are expected for a Milky Way dark matter halo, where the dark matter halo has been constrained with observations of dynamical tracers in the Milky Way halo \citep{battaglia2005,cantena2010,deason2012,nesti2013,kafle2014,zhai2018}. These observations are expected to reflect some degree of contraction of the halo, so we use a publicly available code, CONTRA \citep{gnedin2004}, in order to estimate a non-contracted concentration for the Milky Way. This process is discussed in detail in \autoref{Section:Adiabatic Contraction}. We determine the concentration of the Milky Way to be $c_{\rm NFW}^{\rm MW} = 15.13^{+1.35}_{-2.58}$, which we will use in our modelling.

\subsection{Spin}
\label{sub:spin}

It has long been thought that proto-dark matter haloes acquire angular momentum due to tidal torques from nearby overdensities \citep[e.g.,][]{peebles1969,doroshkevich1970,jones1979,white1978}.
The resulting angular momentum is often parameterized using a dimensionless quantity called the spin parameter. The two most common definitions of the halo spin parameter are  
\begin{equation}
\lambda_{\rm P}  =  \frac{J_{\rm vir}|E|^{1/2}}{GM_{\rm vir}^{5/2}} 
\label{eq:peebles}
\end{equation}
and
\begin{equation}
\lambda_{\rm B} =  \frac{J_{\rm vir}}{\sqrt{2}M_{\rm vir}r_{\rm vir}V_{\rm vir}}, 
\label{eq:bullock}
\end{equation}
where $M_{\rm vir}$, $J_{\rm vir}$, $E$, $V_{\rm vir}$, and $R_{\rm vir}$ are respectively halo virial mass, total angular momentum within the virial radius, total energy of the halo relative to a zero point at infinity, halo virial velocity, and the halo virial radius. The first of these, $\lambda_{\rm P}$, is generally referred to as the Peebles spin parameter and quantifies the angular momentum of the halo in units of the angular momentum necessary to support the halo assuming that all particles are on circular orbits \citep{peebles1980}. The Bullock spin parameter, $\lambda_{\rm B}$, is a convenient definition for cases in which halo energies are not readily available \citep[e.g., in most numerical simulations;][]{bullock2001a}. In this work we use only the Bullock spin parameter. 

Dark matter haloes are mostly supported by the random motions of their particles instead of rotation, so typical values of the spin parameter are quite small, with the median spin value being $\sim 0.05$ and ranging from $0.02-0.11$ \citep{barnes1987}. Since spin characterises the angular momentum of the halo, there is expected to be a correlation between halo spin parameter and galaxy morphological type (with rotationally-supported galaxies found in haloes of greater spin). For example, \citet{vitvitska2002} and \citet{klypin2002} have estimated that Sb galaxies should reside in haloes with spins in the range $\lambda^{\rm Sb}_{\rm P} \sim 0.02$ to $0.10$ (90\% confidence region). Simple galaxy formation and evolution models, such as the classic model of \citet{mmw1998a}, suggest that for a fixed host halo circular velocity, more compact disks form within haloes of lower spin. Therefore, given that the Milky Way has a more concentrated stellar disk than is typical for a galaxy of its mass, we would also expect it to have a lower spin \citep{mm2000}. As an example, the \citet{mmw1998a} model for disk formation in hierarchical cosmologies in particular predicts that 
\begin{equation}
\lambda_{\rm P} = \frac{2.0R_{\rm d}V_{\rm c}(|E|)^{1/2}}{GM_{\rm vir}^{3/2}}\left(\frac{m_{\rm d}}{j_{\rm d}}\right),
\end{equation} 
with the relation from \citet{bullock2001}
\begin{equation}
\lambda_{\rm B} \simeq \lambda_{\rm P}f(c_{v})^{1/2},
\end{equation}
where $j_{\rm d} = \frac{J_{\rm d}}{J}$ is the angular momentum fraction or the disk angular momentum divided by the total angular momentum, $m_{\rm d} = \frac{M_{\rm d}}{M}$ is the disk mass fraction or the disk mass divided by the total mass, $R_{\rm d}$ is the disk scale-length, $V_{\rm c}$ is the circular velocity, and $f(c_{v}) \simeq [2/3+(c_{\rm NFW}/21.5)^{0.7}]$ \citep{mmw1998a,mmw1998b}. For typical concentrations of $c_{\rm NFW} \simeq 10$, $f(c_{v})$ is of order unity. This is what we will adopt for our work.

For simplicity, we use the \citet{mmw1998a} assumption that $\frac{m_{\rm d}}{j_{\rm d}} = 1$ for an NFW halo. This is based upon the assumption that the specific angular momentum of what forms the disk and the halo are the same; however, recent simulations suggest that this idealisation may be suspect \citep{teklu2015,jiang2018}. Recent estimates of the Milky Way parameters include disk scale length ($R_{\rm d} = 2.71^{+0.22}_{-0.20}$ kpc from \citet{licquia2016}, consistent with other measurements \citep[e.g.,][]{mcmillan2011,bland2016}), circular velocity ($V_{\rm c} = 219 \pm 20$ km\,s$^{-1}$ from \citet{reid1999} and \citet{dutton2012}), and virial mass ($M_{\rm vir} = 1.3 \pm 0.3 \times 10^{12} \rm M_{\odot}$). Putting these together, we estimate $\lambda_{\rm B}^{\rm MW} = 0.0332 \pm 0.0105$ for the Milky Way. For comparison, \citet{kafle2014} uses contrasting values for the Milky Way based on kinematics from giant stars, $R_{\rm d} = 4.9 \pm 0.4$ kpc and $M_{\rm vir} = 0.8^{+0.31}_{-0.16}\times10^{12}\rm M_{\odot}$, which yields $\lambda_{\rm B} = 0.0975 \pm 0.0432$ for the Milky Way (or $\lambda_{\rm B} = 0.0596 \pm 0.0193$ if we use the same $M_{\rm vir}$ as in our calculation). In all three cases errors are calculated by propagation of errors.

Our estimate of $\lambda_{\rm B}^{\rm MW} = 0.0332 \pm 0.0105$ 
is consistent with the results of \citet{dutton2012}, 
which use a slightly different approach and slightly different (older) estimates of Milky Way properties. 
This $\lambda_{\rm B}^{\rm MW}$ will be the Milky Way mean and $\sigma$ used in our analysis. We emphasise that we are not looking to calculate a precise $\lambda_{\rm B}$ for the Milky Way - we are more interested in the rank of the Milky Way's spin, and whether it correlates with disk size, as it would in a \citet{mmw1998a} model. Although recent studies have challenged this model, we will take the classic approach. Additionally, as we will show in \autoref{section:satellites}, the spin parameter has a sub-dominant effect on subhalo abundance.

\subsection{Shape}
\label{sub:shape}

Dark matter haloes in $\Lambda$CDM are not spherical but, rather, more nearly triaxial ellipsoids. The shapes of CDM haloes are commonly described by the ratios of their principal axis ratios, $b/a$ (the intermediate-to-long axis ratio) and $c/a$ (the short-to-long axis ratio). Generally, CDM haloes are close to prolate \citep{allgood2006}, with $c \sim b < a$. Therefore, for simplicity we quantify halo shape using $c/a$. Halo shape is highly dependent on the merger history of the halo. The more recent a merger, the less spherical a halo will be and the longest axis of the halo typically correlates with the impact direction of the most recent merger event. Generally, haloes at fixed mass that have formed earlier tend to be more spherical \citep[$c/a \sim 1.0$;][]{schneider2012,white1996}, and more massive haloes tend to be less spherical \citep[$c/a < 1.0$;][]{maccio2007}.

Constraints on the shape of the dark matter halo which hosts the Milky Way are relatively weak. Through various gas and stellar stream measurements, density profile estimates, and simulations, the Milky Way is approximated to be quasi-spherical, with $c/a^{\rm MW} \simeq$ 0.72--0.8 according to \citet{law2010} and \citet{veraciro2013} when incorporating measurements from the Sagittarius stream. Therefore the estimated value for the Milky Way we use in our analysis is $c/a = 0.76$, $\sigma_{c/a} = .02$.

\subsection{Halo Merger History}

Halo merger history is also correlated with subhalo abundance. Earlier forming haloes have been shown to have less substructure \citep{zentner2005,jiang2016,mao2018}. Host haloes that assembled earlier are expected to end up with less mass in subhaloes because there has been more time for accretion by the host.


There is good evidence from chemo-dynamical studies \citep[e.g.,][]{ruchti2015} and other work that the Milky Way has had a quieter accretion history than typical. It appears that the Galaxy has not had any substantial mergers since the formation of its galactic disk \citep[$\approx$ 9--12 Gyr ago;][]{unavane1996}. 

For simplicity, we characterise merger history in terms of the scale factor of the Universe at the time of a halo's last major merger, $a_{\rm LMM} = \frac{1}{1 + z_{\rm LMM}}$. For the zoom-in simulations used in this paper, we define a major merger as one with a mass ratio $> 0.3$ (which implies that a merger like the current Milky Way--Sagittarius merger would be classified as a minor merger, as expected by current mass estimates). Using the limits of 9--12 Gyr for the Milky Way's last major merger we convert to $z = 1.33 - 3.55$ and $a_{\rm LMM}^{\rm MW} = 0.43 - 0.22$ respectively of which we select the median $a_{\rm LMM}^{\rm MW} = 0.325$), using WMAP9 cosmological parameters (which match the simulation parameters closely; cf. \autoref{sub:simulations}). The choice of cosmology has a minuscule effect on this estimate in comparison to the uncertainty in the time since the last major merger. 

\subsection{The Milky Way Halo Compared to Other Dark Matter haloes}
\label{sub:comparison}

\begin{figure*}
\includegraphics[width=\linewidth]{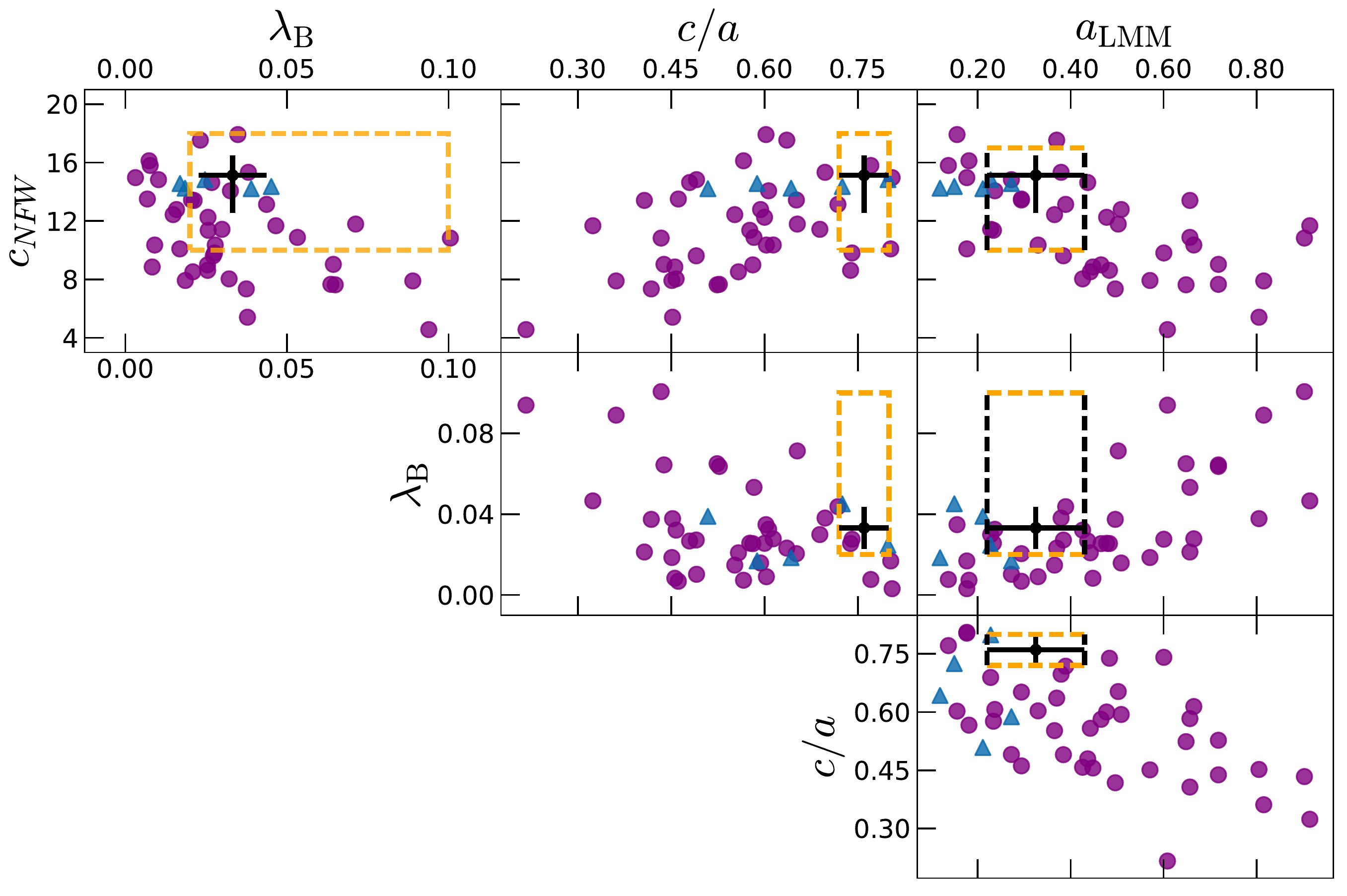} 
\caption{Plots of all possible combinations of the host halo parameters investigated in this paper. The orange dashed region indicates the range of estimated values for Sb galaxy host haloes; these have morphology similar to the Milky Way (which is estimated to be SBb/c). The black dashed regions shown in $a_{\rm LMM}$ are values quoted only for the Milky Way due to no known work on major merger scales of Sb galaxy dark matter haloes. 
The black point with errorbars represent the estimated properties for the Milky Way host halo described in \autoref{Section:milkyway}. We emphasise that these are approximate estimates and in many cases are difficult to constrain well. The blue triangle points are the 5 nearest neighbours to the Milky Way within this four-dimensional parameter space; their identification is discussed in \autoref{sub:neighbors}. The Spearman correlation coefficient and $p$-values for all of the correlations shown here are presented in \autoref{tab:spearman}. The Milky Way lies at the outskirts of each of these projections. Known relationships between host halo properties have been reproduced by our simulations.}
\label{fig:triangle}
\end{figure*}

Figure \ref{fig:triangle} depicts the joint distributions of the spin ($\lambda_{\rm B}$), concentration ($c_{\rm NFW}$), shape ($c/a$), and last major merger scale ($a_{\rm LMM}$) parameters of the 45 Milky Way-size halo simulations that we study. Each host halo from a zoom-in simulation is depicted as a circular or triangular point in purple or blue respectively. The blue triangular haloes correspond to the five nearest neighbours to the Milky Way in the multi-dimensional space consisting of all the parameters plotted, as we will discuss in \autoref{sub:neighbors}. Our simulated Milky Way-like haloes exhibit the same correlations between these properties found in prior work. For example concentration and spin are known to be anti-correlated \citep{maccio2007}, and shape and merger history are expected to be correlated \citep{allgood2006}.

The region of \autoref{fig:triangle} in which Sb galaxies' haloes are thought to reside is shown by the orange dashed region. We highlight this regime to enable comparisons to 
the estimated parameters of the Milky Way halo, since the Milky Way is generally classified as an SBb/c galaxy. In the case of $a_{\rm LMM}$, a value is quoted only for our Galaxy, denoted by the black dashed lines, as we are not aware of prior work on major merger scale parameters for Sb galaxy dark matter haloes. 

The black points with the error bars represent the estimated parameters of the Milky Way host halo  from the literature; their provenance is described individually above. As is evident, the Milky Way lies closer to the outskirts of the multidimensional distribution in each projection. In particular, the Milky Way halo appears to have a somewhat low spin, high concentration, more spherical shape, and a longer lookback time to the last major merger than a typical halo of the same mass. The concentration, shape, and lookback time of the Milky Way is expected to be consistent with the Galaxy's more compact stellar disk than average at fixed luminosity.

We have demonstrated that the Milky Way may well be an outlier in the distributions of several halo properties. One of these properties, concentration, was previously found to correlate with subhalo abundance \citep{zentner2005,mao2015,jiang2016}. Indeed, it is not unreasonable to suspect that other properties correlate with subhalo abundance as well, particularly because halo spin and shape are so strongly associated with halo merger activity. In the following section, we will investigate the correlations between host halo properties and subhalo abundance and use these correlations to make predictions for subhalo demographics within haloes resembling that of the Milky Way.

\section{Subhalo Abundances in Milky Way-Like haloes}
\label{section:satellites}
In this section we investigate the relationships between the host halo properties described above and the abundance of subhaloes in each simulated Milky Way-like halo. We will then incorporate the correlations observed into a prediction for subhalo abundances within haloes resembling that in which the Milky Way resides.

\subsection{Halo Properties and Subhalo Abundances}
\label{sub:props_and_abundance}
We begin with a simple statistical search for correlations. \autoref{tab:spearman} shows the Spearman correlation coefficient between the host halo properties. The Spearman (or ranked) correlation coefficient ($\rho$)  measures the strength and direction of a monotonic relationship between two ranked variables \citep{spearman}. The coefficient can range from $+1$ to $-1$, where the extremes indicate that each of the variables is a strictly monotone function of the other, so that that ranks within lists of the two variables are perfectly associated. The $p$ values is a way of investigating whether we can accept or reject the null hypothesis that there is no monotonic association between the two variables. We set our threshold at $p < 0.05$, or a less than 5\% chance that the relationship found (or any stronger relationship) would happen if the null hypothesis were true. In \autoref{tab:spearman} the numbers above the diagonal denote $\rho$ and the numbers below the diagonal denote $p$. The table is colour-coded according to the correlation coefficient: values of $\rho$ near 1.0 are shown as red while $\rho$ near $-1.0$ corresponds to blue. The Spearman correlation is sensitive to both linear and non-linear relationships, and (unlike the Pearson correlation coefficient) is robust to outliers. We use the Spearman correlation because we are interested in testing for general monotonic relationships between dark matter halo properties.

Focusing on the first column and row, we conclude that $c_{\rm NFW}$, $a_{\rm LMM}$, and $\lambda_{\rm B}$ are all significantly correlated with subhalo abundance. The shape parameter $c/a$ still has a relationship with subhalo abundance, but not as strong as for the other host properties.

Although no significant correlation between the abundance of subhaloes and host halo mass is found here, this is almost certainly due to the small mass range of the zoom-in haloes, such that the variations in other parameters dominate.  In the parent simulations from which the re-simulated haloes were drawn, the number of subhaloes is on average directly proportional to host halo mass.

\begin{table*}
\centering\includegraphics[width=1.0\textwidth]{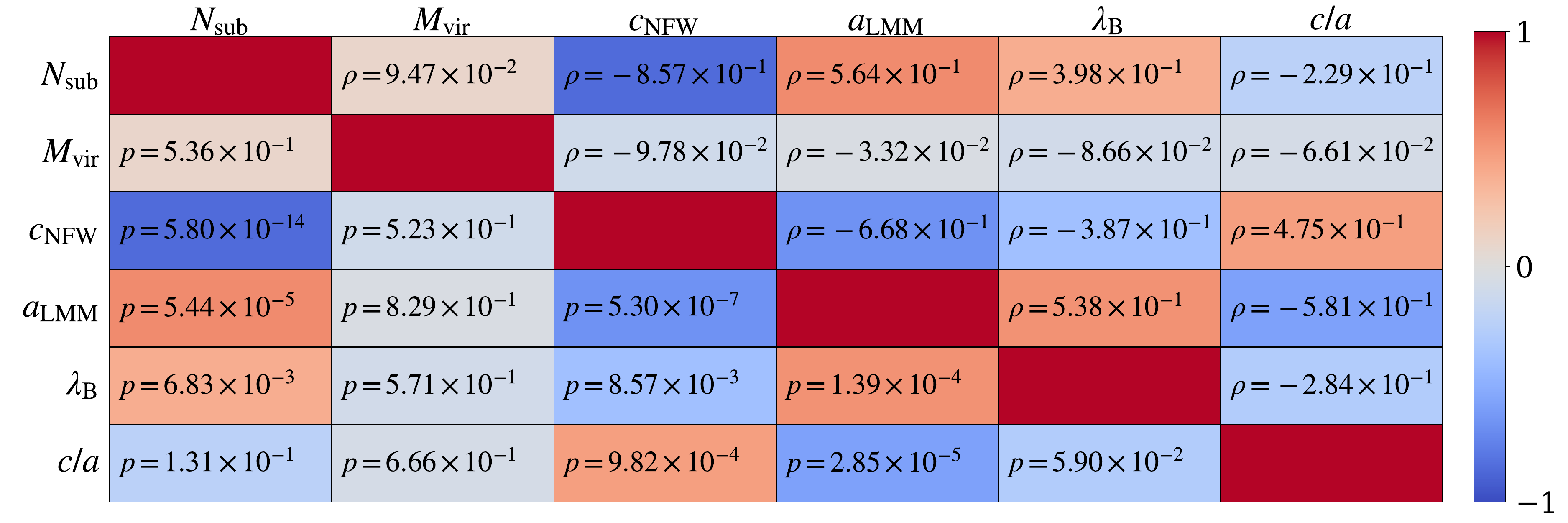}
\caption{Table of Spearman correlation coefficients, $\rho$, and the corresponding $p$-values, $p$, amongst all of the host halo properties considered in our analysis: the number of subhaloes ($N_{\rm sub}$), the mass within the virial radius ($M_{\rm vir}$), concentration ($c_{\textrm{NFW}}$), the scale factor at the time of the last major merger ($\alpha_{\rm LMM}=\frac{1}{1+z_{\rm LMM}}$), the Bullock spin parameter ($\lambda_{\rm B}$), and the halo shape (parameterised by the ratio of tertiary to major axis length, $c/a$). Above the diagonal and denoted by $\rho$ are the correlation coefficients. Below the diagonal and denoted by $p$ are the corresponding $p$-values for each correlation. A value $p<0.05$ corresponds to a correlation that is statistically significant at $>2\sigma$, or a $ < 5\%$ chance this relationship would be found if the null hypothesis of of no monotonic association between variables is true. The values are coloured by the strength of their correlation: 1.0 is red and -1.0 is blue, as described in the colourbar. Subhalo number is most strongly correlated with $c_{\textrm{NFW}}$, $\alpha_{\rm LMM}$, and $\lambda_{\rm B}$ in our simulations, followed by $c/a$ and very weakly by $M_{\rm vir}$ (which is expected given our small mass range).}
\label{tab:spearman}
\end{table*}

\begin{figure*}
    \centering
    \begin{minipage}{0.25\textwidth}\centering
        \includegraphics[width=\textwidth]{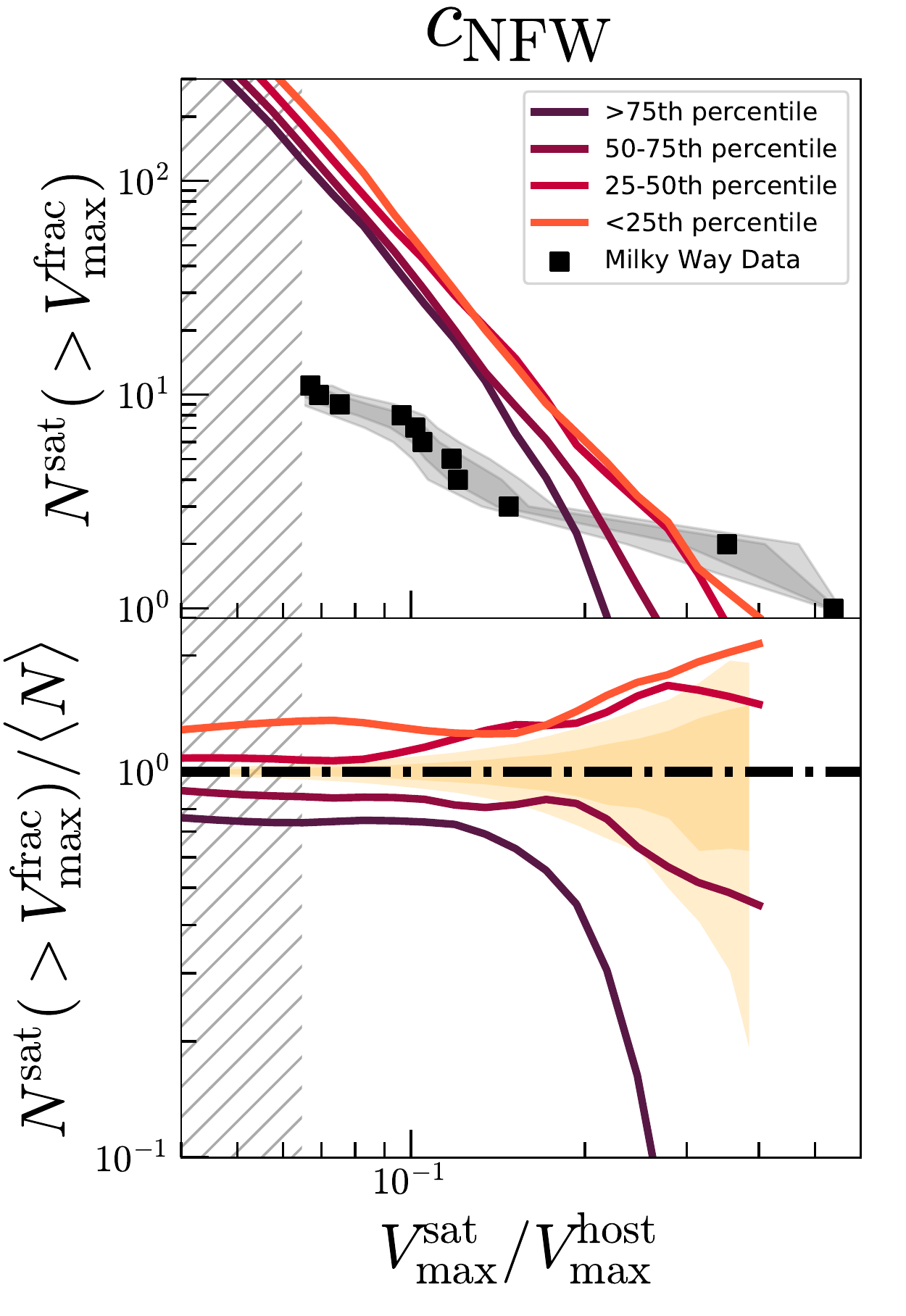}\\ (a)
    \end{minipage}%
    \begin{minipage}{0.25\textwidth}\centering
        \includegraphics[width=\textwidth]{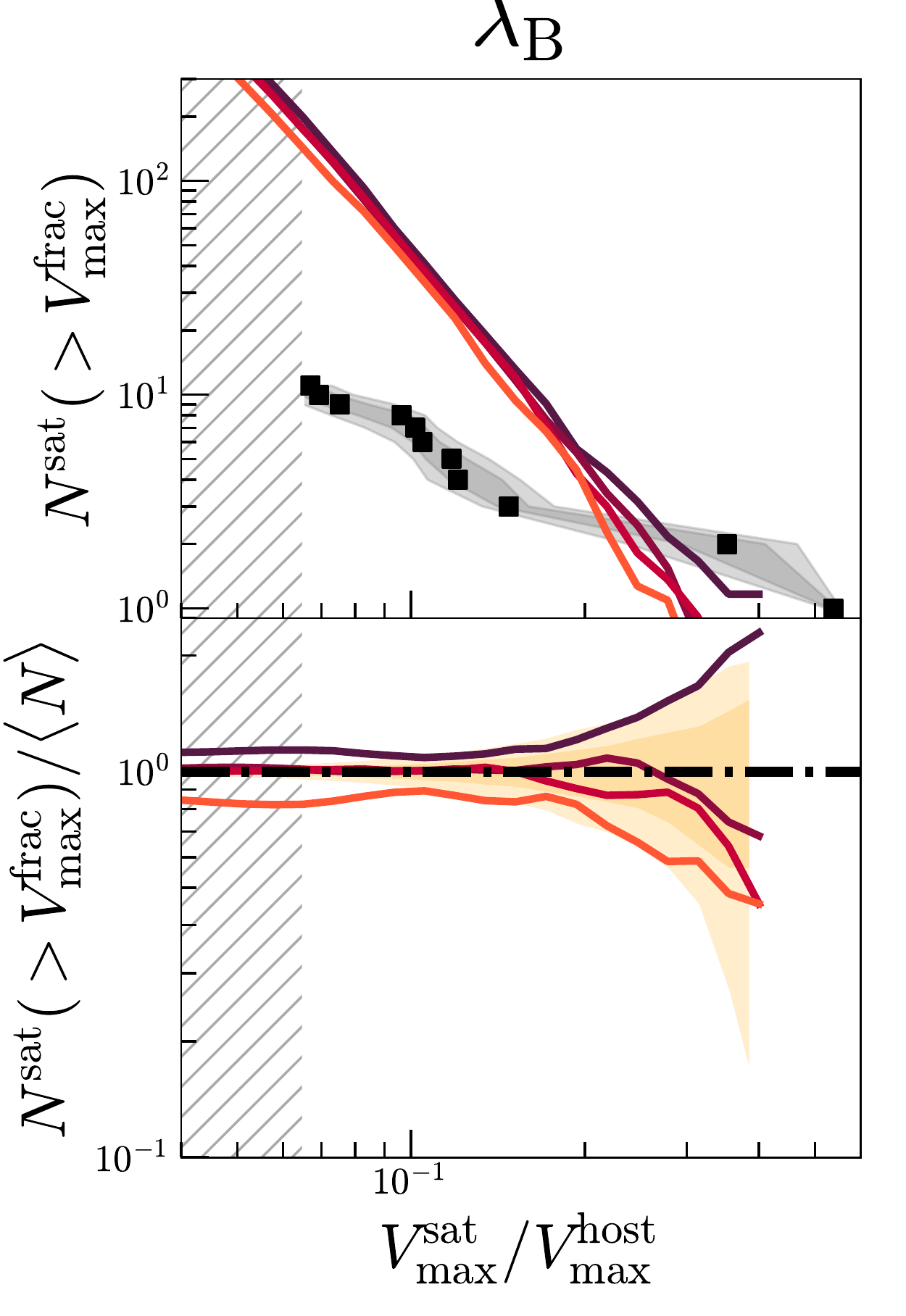}\\ (b)
    \end{minipage}%
    \begin{minipage}{0.25\textwidth}\centering
        \includegraphics[width=\textwidth]{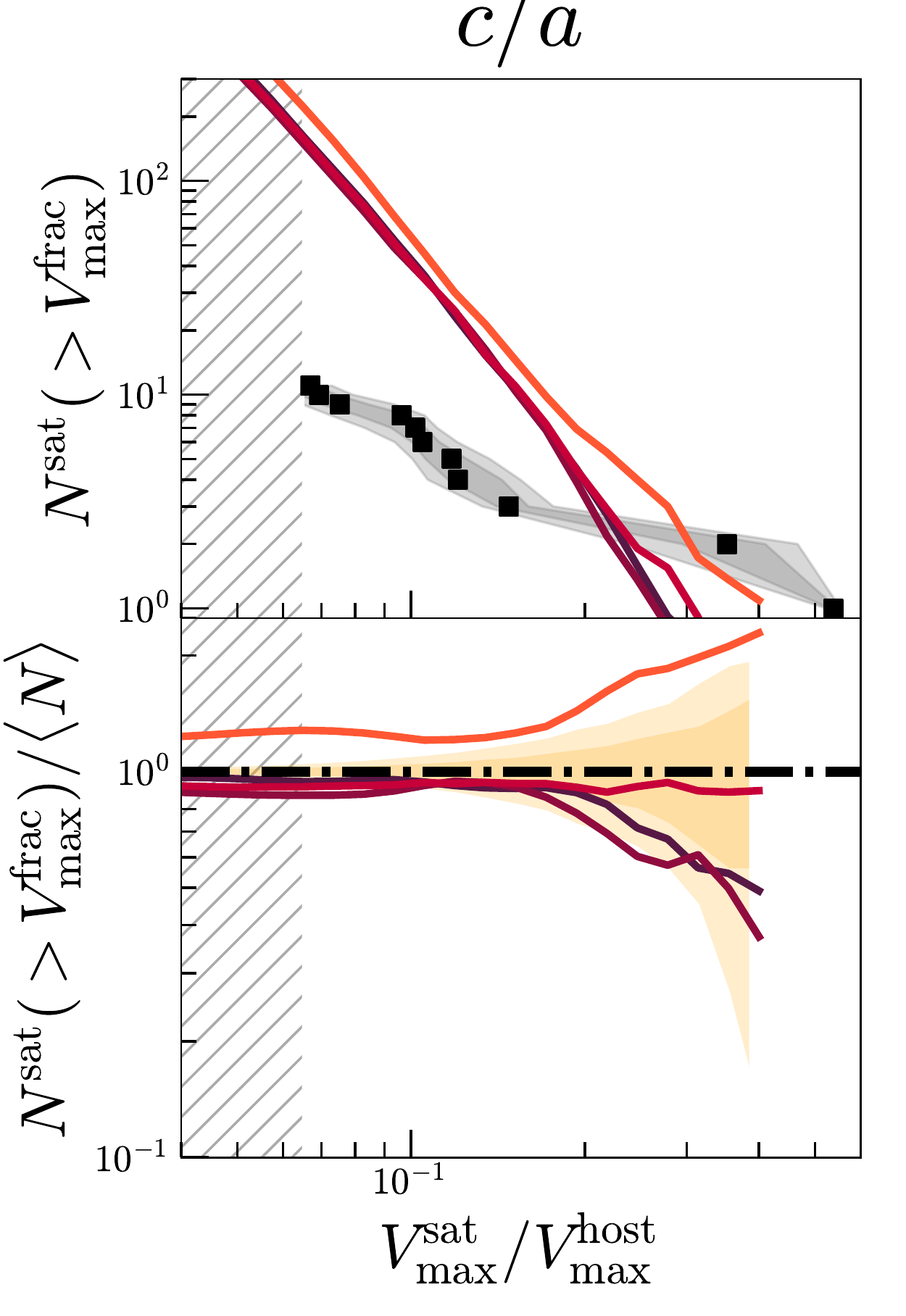}\\ (c)
    \end{minipage}%
    \begin{minipage}{0.25\textwidth}\centering
        \includegraphics[width=\textwidth]{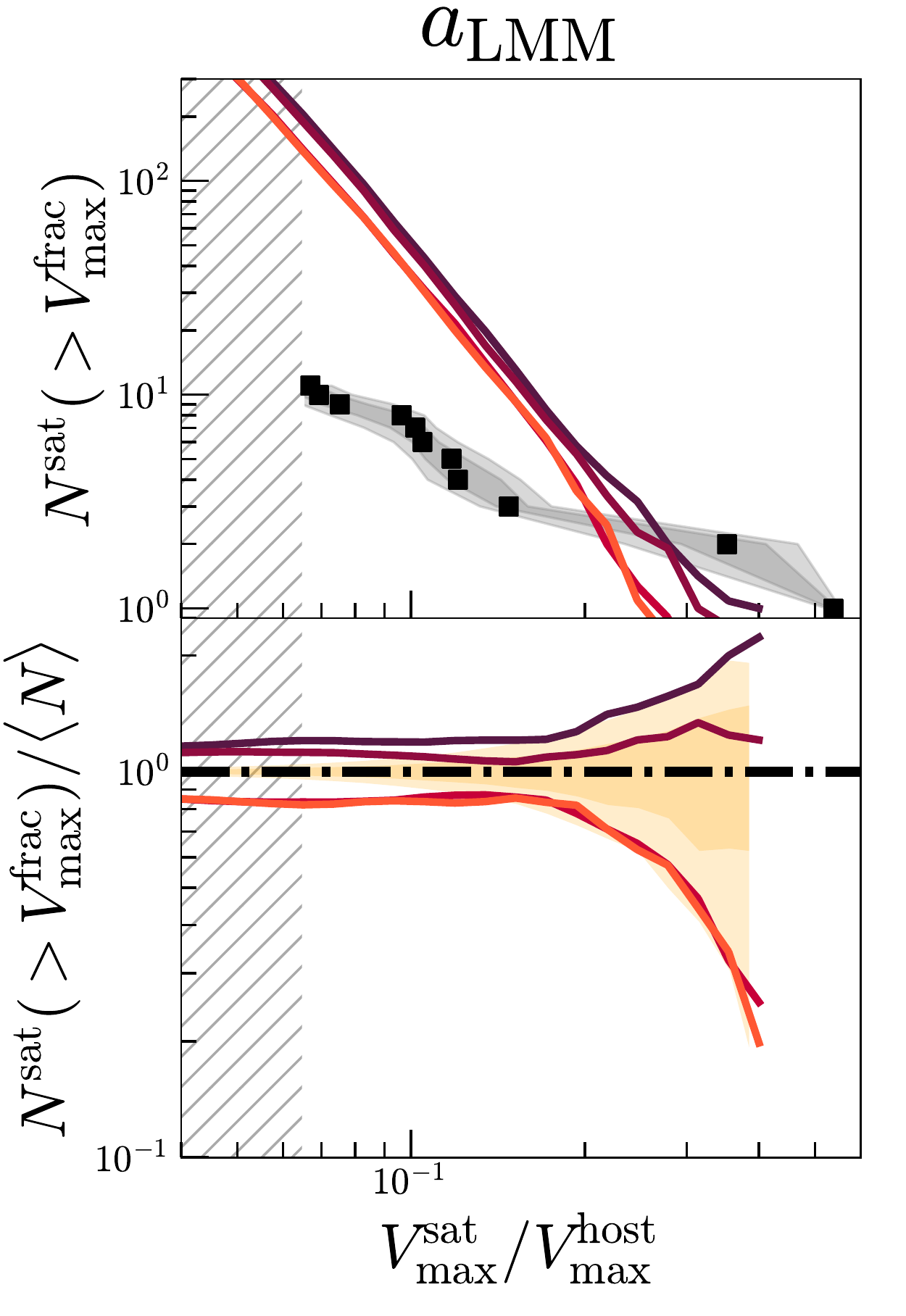}\\ (d)
    \end{minipage}%
\caption{Cumulative velocity functions $N_{\rm sat}(>V_{\rm max}^{\rm sat}/V_{\rm max}^{\rm host}=V_{\rm max}^{\rm frac})$ of subhaloes for samples split according to various host halo properties.  The black points indicate the 11 classical Milky Way satellites ($M_{\rm vir}\simeq 10^{10}\rm M_{\odot}$). The grey region around each point represent 68\% and 95\% confidence regions as a result of Gaussian perturbations with amplitude given by the error in each satellite's $V_{\rm max}^{\rm frac}$. Separate cumulative velocity functions are shown for haloes divided into quartiles by, from left to right: (a) concentration ($c_{\rm NFW}$), (b) spin ($\lambda_{\rm B}$), (c) shape ($c/a$), and (d) scale factor at last major merger ($a_{\rm LMM}$); colour labelling is indicated in each panel's legend. In the bottom panel of each plot is shown the ratio of each curve to the average cumulative velocity function including all host haloes.  These normalised plots include a black dashed line at $N_{\rm frac}(>V_{\rm max})/\langle N\rangle = 1$ to indicate where there is no separation between the quartiles. The orange regions around $N_{\rm sat}(>V_{\rm max}^{\rm frac})/\langle N\rangle = 1$ indicate the 68\% and 95\% confidence regions for a case with no deviation from the mean value due to Poisson errors (given the average number of subhaloes and the total number of host haloes that fall in a quartile where we use 11). Portions of the figures below the resolution limit of the simulations used, $V^{\rm sat}_{\rm max}/V^{\rm host}_{\rm max} = 0.065$  (described in \autoref{sub:simulations}), are depicted by the hatched regions. At low velocities the cumulative velocity functions exhibit statistically significant separations amongst various host halo properties.}
\label{fig:CVFs}
\end{figure*}

We next investigate \textit{how} each host property individually influences subhalo abundance. In \autoref{fig:CVFs} we present the mean cumulative velocity function (CVF) of the subhaloes when host haloes are divided into subsets according to their properties. The CVFs (top panels) and ratio of each CVF to the average halo CVF (bottom panels) are separated into quartiles according to each host halo property considered here: $c_{\rm NFW}$, $\lambda_{\rm B}$, $c/a$, or $a_{\rm LMM}$. The vertical axis of the CVF (upper) plots is the average cumulative number of subhaloes above a threshold in velocity, while the horizontal axis is the corresponding maximum subhalo velocity normalised by the maximum velocity of the respective host. The vertical axis in the lower panel corresponds to the cumulative number of subhaloes for a particular quartile divided by the average cumulative number of subhaloes amongst all hosts. 

The quartile curves shown in each panel are determined by percentiles in each respective host halo property. The $>75^{\rm th}$ percentile bin for each panel includes 12 host haloes, while the other percentiles have 11 host haloes each. For each halo we determine how many subhaloes are in each of 20 logarithmically spaced $V_{\rm max}^{\rm sat}/V_{\rm max}^{\rm host} = V_{\rm max}^{\rm frac}$ bins; from this we can determine the cumulative number of subhaloes for each halo summing down to a given bin of $V_{\rm max}^{\rm frac}$. The mean CVF for the haloes in each quartile are shown as the red to orange lines. Portions of the plots which lie below the resolution limit described in \autoref{sub:simulations} (i.e., with $V^{\rm frac}_{\rm max} < 0.065$) are indicated by the hatched region. 

The black points represent the 11 classical satellites of the Milky Way, using values from \citet{xue2008, vanderMarel2014,lallivayalil2013, kuhlen2010,b-k2012,rashkov2012,mcc2012} and \citet{jiang2015} compiled in Table 1 of \citet{jiang2015}. The only kinematic information available for the Milky Way dwarf spheroidals is the line-of-sight velocities of stars, which can be used to constrain the dynamical mass of the dwarf. In the case of Sculptor, Draco, Leo II, Fornax, Sextans, Carina, Leo I, and Ursa Minor, \citet{kuhlen2010} and \citet{b-k2012} use the Via Lactea II simulation or the Aquarius suite of simulations, respectively, to assign weights to subhaloes in the simulations according to how well they match the dynamical mass of each respective Milky Way satellite, and then use the weighted average of $V_{\rm max}$ for those subhaloes as an estimate of the satellite's $V_{\rm max}$ value. For example, \citet{b-k2012} computes a distribution function of possible $V_{\rm max}$ by assigning a weight from the estimated likelihood that each subhalo from their six randomly-selected Milky Way-mass host haloes is consistent with the given satellite's mass. For the case of the Large Magellanic Cloud (LMC) \citet{vanderMarel2014} uses proper motions and line of sight velocity measurements of stars in the LMC in concordance with a model of a flat rotating disk to estimate the circular velocity. A similar treatment is done for the Small Magellanic Cloud (SMC). In the case of Sagittarius, \citet{jiang2015} uses the relation $V_{\rm max} = 2.2\sigma_{\rm line\;of\;sight}$ \citep{rashkov2012} with the line of sight velocity dispersion measurement ($\sigma_{\rm line\;of\;sight}$) from \citet{mcc2012} in order to estimate its $V_{\rm max}$. The dwarf spheroidal estimates from \citet{kuhlen2010,b-k2012} and \citet{vanderMarel2014} are all consistent (within errors) with this relation.

A key assumption made is that the simulated haloes of a given mass will match the kinematics of Milky Way satellites' haloes of the same estimated mass. In particular, the stellar content of these satellites is only in the very central region ($\sim 1$ kpc) so extrapolation beyond this stellar distribution is necessary to constrain $V_{\rm max}$. Although using $V_{\rm max}$ is less subject to extrapolating issues than the total mass, $V_{\rm max}$ will still depend substantially on the assumed distribution of dark matter for a galaxy of the observed size \citep{zentner2003}. The shape of the dark matter density profile  may vary substantially from an NFW profile. In the case for the work by \citet{b-k2012}, the density profiles are not assumed NFW or Einasto with all properties computed from the raw particle data in order to get around this issue (but a general profile is still indeed assumed).

To normalise these values to $V_{\rm max}^{\rm frac}$ we use the average $V_{\rm max}$ of the Milky Way after doing 10,000 bootstraps perturbing the measured value from \citet{xue2008} and \citet{jiang2015} of $V_{\rm max}^{\rm MW} = 170 \pm 15$ km\,s$^{-1}$ by a Gaussian of the error. This value was determined by using line of sight kinematic data and connecting it with simulation data by finding the best matched probability distributions. Our resulting $V_{\rm max}^{\rm host} = 170.22$ km\,s$^{-1}$ for the Milky Way. This is consistent with our Milky Way-mass host haloes, that have an average $V_{\rm max}^{\rm host} = 174.06$ km\,s$^{-1}$.

The shaded grey regions around the Milky Way satellite points indicate the $68\%$ and $95\%$ confidence regions from the effect of measurement errors on each satellite $V_{\rm max}$. For each satellite, we generate 10,000 Gaussian-distributed values randomly drawn from the errors in each $V_{\rm max}^{\rm sat}$ and then perturb the estimated value for that satellite by the generated value. We emphasise that the goal of including the Milky Way satellite points is strictly for reference and not direct comparison, as our dark matter-only simulations do not include the baryonic physics, feedback mechanisms, etc.\ (see \autoref{Section:intro}) that would be necessary to make the $V^{\rm frac}_{\rm max}$ values from the simulations directly comparable to the Milky Way satellite characteristics. 

The bottom panel of each plot, which depicts the ratio of each quartile's mean CVF to the overall average CVF, shows the differences amongst the quartiles for a given property more clearly than the raw cumulative velocity functions plotted in the top panel. A dotted horizontal line at $N_{\rm sat}(>V_{\rm max}^{\rm sat})/\langle N\rangle = 1$ indicates where there would be no difference between a given quartile and the mean. The orange regions around $N_{\rm sat}(>V_{\rm max}^{\rm sat})/\langle N\rangle = 1$ corresponds to the $68\%$ and $95\%$ confidence region about this value from Poisson errors for a quartile of 11 haloes.

The differences between the CVFs of quartiles divided according to a given property allow us to investigate the relationship between that property and subhalo abundance. In \autoref{fig:CVFs} it is clear that at low velocities, the separation between the extreme quartiles for every property shown is larger than the 2$\sigma$ Poisson error. We see the most significant separation when we divide samples according to $c_{\rm NFW}$; this property also has the strongest correlation to $N^{\rm sub}$ (cf. \autoref{tab:spearman}), consistent with the results from \citet{zentner2005} and the model developed by \citet{mao2015}. The next strongest effect is associated with $a_{\rm LMM}$, as expected from predictions from e.g., \citet{zentner2005} and \citet{jiang2016}. Interestingly $c/a$ shows a more significant separation than $\lambda_{\rm B}$ in \autoref{fig:CVFs}, in contrast to \autoref{tab:spearman}. The higher concentration, lower spin, more spherical, or earlier forming haloes -- that is, those which are most similar to the estimated properties of the Milky Way dark matter halo in each characteristic -- are all associated with having fewer subhaloes. 

We can conclude that at low velocities this set of four host halo properties can help to predict subhalo abundance, given their correlations with that quantity. We expect this to be the case at higher velocities as well, but there is too much noise due to low counts per bin to draw a statistically significant conclusion from \autoref{fig:CVFs} at high velocities. Physically we expect there to be far more subhaloes with low $V_{\rm max}$ than high, given the mass function of subhaloes \citep[e.g.,][]{moore1999,bullock2000,stoehr2002,kravtsov2006}, making it the most important region to probe. 

Having shown that we can identify host halo properties that correlate with subhalo abundance, we next investigate what \textit{combinations} of these parameters provides the best predictions of subhalo abundances for Milky Way-like dark matter haloes. 

\subsection{Predicting Milky Way Subhalo Abundances}
\label{sub:mw_model}

To address this question, we have built power-law scaling relation models which utilise various combinations of halo properties as predictors for the cumulative number of subhaloes above a given value of $V_{\rm max}$, in order to produce more accurate predictions of the subhalo abundance for the Milky Way. We describe these models in detail in \autoref{sub:models} but summarise them here.

The first model considered is a relatively simple one, incorporating only $c_{\rm NFW}$ 
to predict subhalo abundance; we refer to it as our ``one-parameter model'' hereafter 
(the fit does incorporate a second parameter setting the scale of the overall subhalo numbers at a given velocity, however). 
This approach can be motivated by \citet{mao2015}'s conclusion that halo concentration provides sufficient information to predict subhalo abundance in haloes of a given mass. We compare predictions from this simple model to results from a power-law model built using an optimised combination of the examined host halo properties, which has greater statistical explanatory power; we will refer to it as our ``three-parameter model'', though again it also incorporates a normalisation factor. We also compare to a model that does not have the best statistical explanatory power, 
but does not include concentration.

Specifically, the robust three-parameter model includes concentration, spin, and shape (as well as the assumption, motivated by tests with larger simulations, that subhalo abundance is proportional to mass). This specific set of parameters was chosen because it had lower Akaike and Bayesian Information criteria (AIC and BIC) than other models considered, which included all combinations of the halo parameters used in this paper; quadratic terms combining those parameters; and first order polynomial cross terms e.g., $c_{\rm NFW}\times c/a$, that had as many as 5 total parameters (apart from a constant term). These low information criterion values indicate that this model provides a better fit for subhalo abundances, \textit{given the number of free parameters in the model}, than any others considered. Details of this evaluation are given in \autoref{sub:models}. This model is expected to provide additional information as opposed to just additional degrees of freedom, given its low AIC and BIC values. Although both models provide useful predictions of subhalo abundances, we would expect the three-parameter model to always provide a more accurate prediction than the one-parameter model, as the one-parameter model is a special case of the three-parameter model (so further optimisation via the other parameters can only improve performance; see \autoref{table:AICBIC} and discussion). We also show in \autoref{table:AICBIC} that the $\chi ^{2}$ value of three-parameter model is $20\%$ better than for the one-parameter model. 

The comparison three-parameter model includes spin, shape, and scale factor at the time of the last major merger. This model is chosen for comparison due to the dubious nature of the concentration measurement of the Milky Way, as well as a means for comparing to a model without concentration. This model's information criteria are also shown in \autoref{table:AICBIC}, which makes it evident that this model is not the best combination of parameters for predicting subhalo abundance. We use these three models to estimate subhalo abundances for the Milky Way; by comparing their results we can assess the robustness of our predictions.

We obtain the best predictions of subhalo abundances with a model where the average number of subhaloes in a halo of given properties has a power-law dependence on all relevant parameters, as described in \autoref{sub:models} and \autoref{Section:maxlikelihood}. We define such a power-law model as:
\begin{equation}
N_{\rm sub}^{\rm pred}(>V_{\rm max}^{\rm frac}) = k\times\prod_{i}x_{i}^{\alpha_{i}}, 
\label{equation:power1}
\end{equation}
where $N_{\rm sub}^{\rm pred}$ is the predicted average subhalo abundance based on halo properties, $k$ sets the scale of the abundances, and $\alpha_{i}$ is the exponent for the $i^{\rm th}$ halo parameter $x_i$ used in the model  (e.g., $c_{\rm NFW}$). 

Because we are trying to predict the {cumulative} subhalo abundance for the Milky Way even at relatively high velocity thresholds where most haloes have few subhaloes, the Gaussian assumption which underlies the method of least-squares linear regression is not valid for this problem (following the usual rule of thumb that the Poisson distribution can be safely approximated by a Gaussian only for $N>25$). Instead, we rely on a Poisson maximum likelihood method to fit models, as it should provide accurate results even in this regime. Specifically, we determine the parameter values which maximise the likelihood of the observed set of subhaloes in the simulations. Given the properties for the Milky Way discussed in \autoref{Section:milkyway} in combination with the results from the maximum likelihood fits, we make a prediction for $N^{\rm sub}(>V_{\rm max}^{\rm frac})$ for the Milky Way in 20 separate bins of $V_{\rm max}^{\rm frac}$, i.e. we fit a separate model for each threshold of $V_{\rm max}^{\rm frac}$.  The equations and algorithms underlying our methods are discussed in detail in \autoref{Section:maxlikelihood}.

For each bin in $V_{\rm max}^{\rm frac}$ we can predict a cumulative number of subhaloes for the Milky Way down to that velocity threshold by substituting in the estimates of the Galaxy's parameters discussed in \autoref{Section:milkyway} for the $x_{i}$ in \autoref{equation:power1}, and using the $k$ and $\alpha_i$ values resulting from the model fit for that bin. For example, in the case of the one-parameter model we use the estimate of $c_{\rm NFW} = 15.13$ for the Milky Way, and the $k$ and $\alpha_{c_{\rm NFW}}$ that result from the Poisson maximum likelihood fit for a particular velocity threshold to obtain a prediction for the corresponding element of the Milky Way CVF.

\begin{figure}
\includegraphics[width=1.0\linewidth]{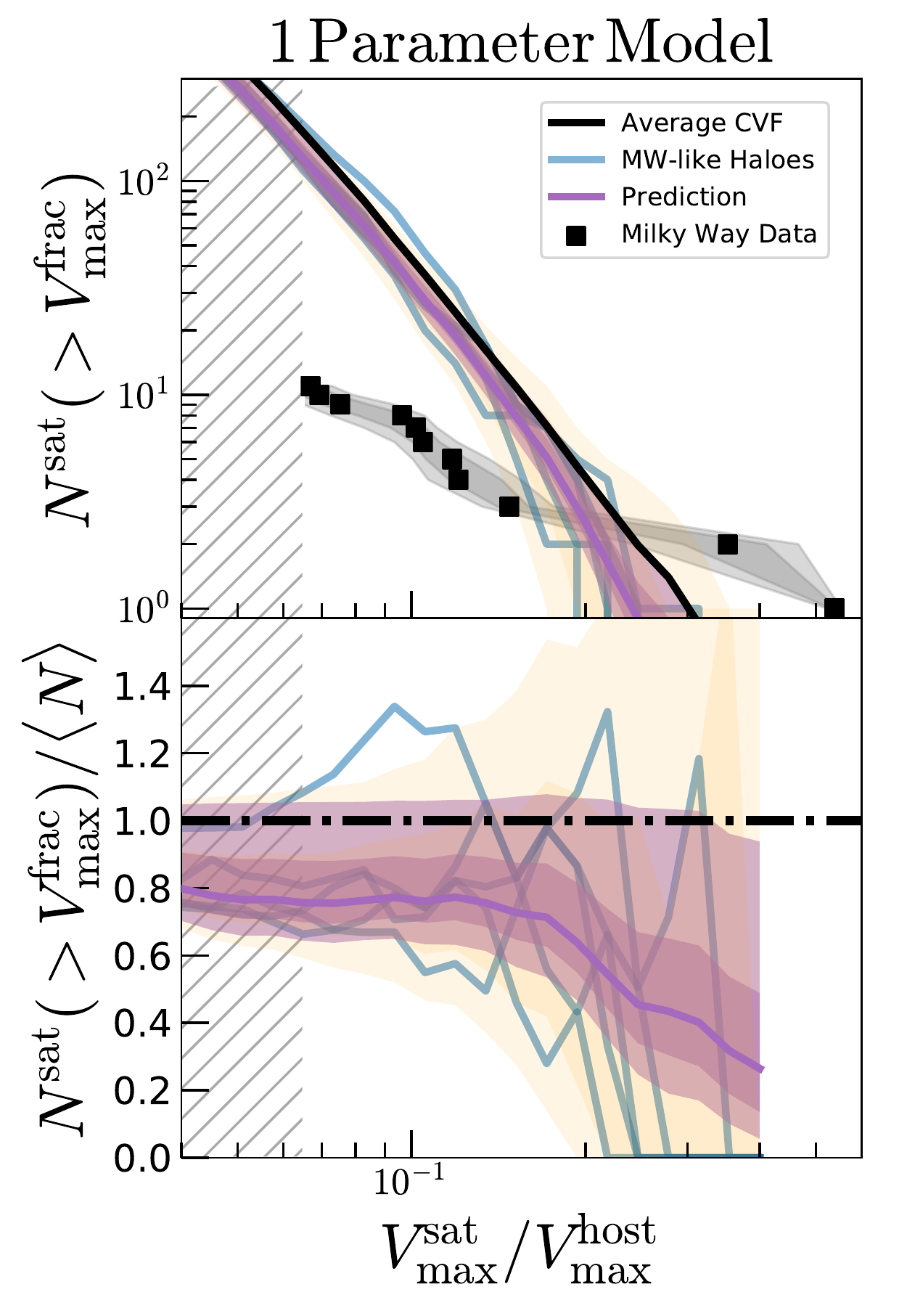} 
\caption{\textit{Top panel}: Cumulative velocity function for the Milky Way from a one-parameter scaling relation model which predicts subhalo abundance as a function of halo concentration. 
The purple line corresponds to the prediction of the model at each threshold value of $V_{\rm max}^{\rm frac}$ for the Milky Way, based on the MW halo's estimated properties. The solid black line corresponds to the average CVF for all of the 45 zoom-in host haloes, while the blue lines are the CVFs for the five nearest neighbours to the Milky Way in parameter space, as indicated in \autoref{fig:triangle} and described in \autoref{sub:neighbors}. The darker and lighter purple regions indicate the $68\%$ and $95\%$ confidence regions about the purple prediction line due to uncertainties in both fit parameters (evaluated via bootstrap re-sampling) and in Milky Way properties (evaluated by re-drawing values from Milky Way parameter uncertainties before evaluating the models). The darker and lighter orange regions indicate $68\%$ and $95\%$ confidence regions which incorporate Poisson scatter in subhalo abundances as well. \textit{Bottom panel}: Ratios of the CVFs to the average CVF of all simulated Milky Way-like haloes. The one-parameter model predicts that the Milky Way's host halo should have (at 68\% confidence) 12--30\% fewer subhaloes than average at the low $V_{\rm max}^{\rm frac}$ end and 19--45\% fewer subhaloes than average at the high $V_{\rm max}^{\rm frac}$ end (though Poisson scatter can dwarf this effect, especially at high $V_{\rm max}^{\rm frac}$) compared to an average dark matter halo of the same mass.}
\label{fig:CVF1param}
\end{figure}

\begin{figure}
\includegraphics[width=1.0\linewidth]{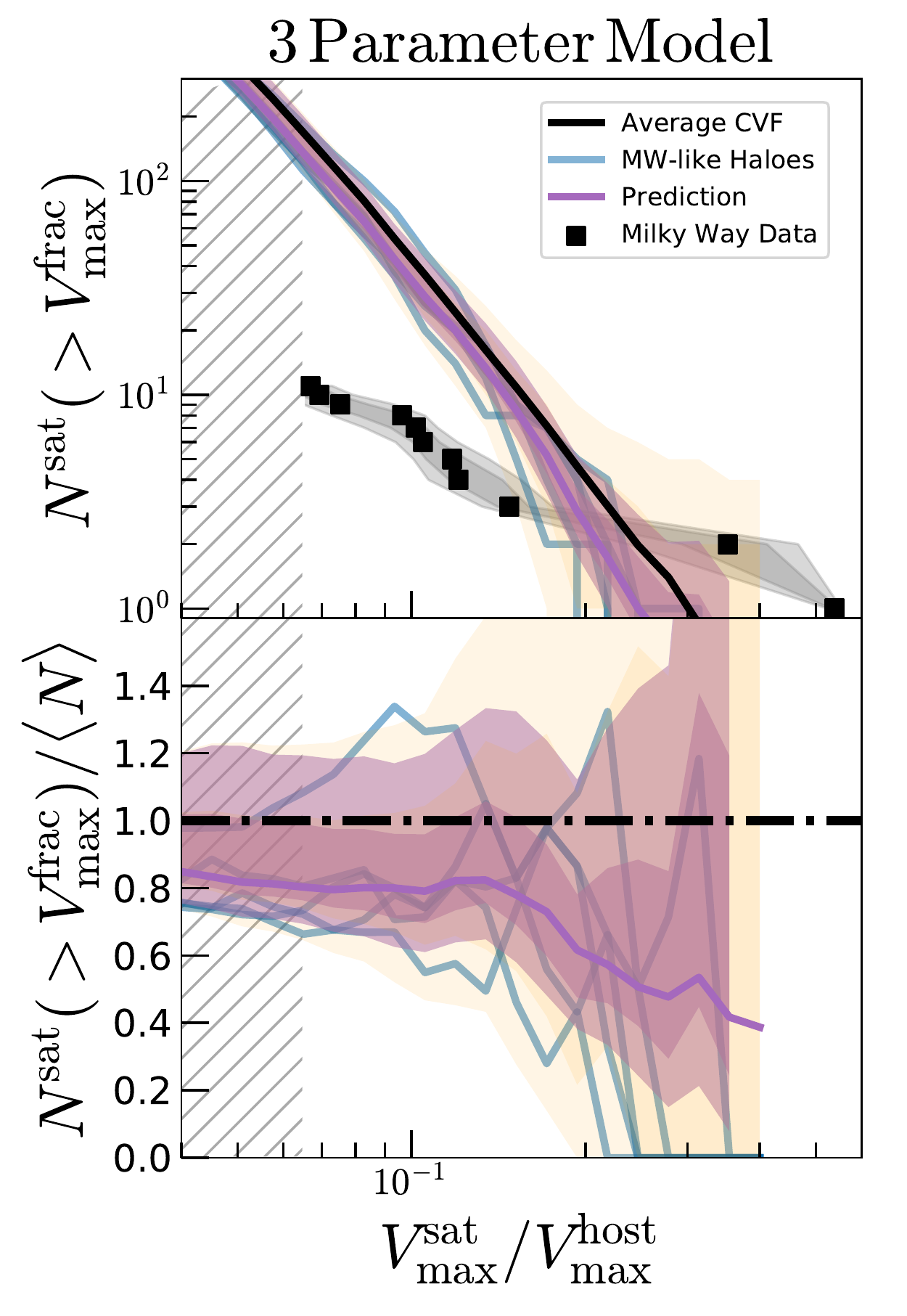}
\caption{As \autoref{fig:CVF1param} but for a three-parameter scaling relation model,  which predicts subhalo abundance as a function of concentration, spin, and halo shape. The three-parameter model has a less significant prediction than the one-parameter model due to the buildup of error. This model predicts that the Milky Way's host halo should have 1--24\% fewer subhaloes than average at the low $V_{\rm max}^{\rm frac}$ end (at 68\% confidence) and 22--52\% fewer subhaloes at the high $V_{\rm max}^{\rm frac}$ end. Again, additional error from Poisson scatter can dwarf this effect, especially at high $V_{\rm max}^{\rm frac}$.}
\label{fig:CVF3param}
\end{figure}

\begin{figure}
\includegraphics[width=1.0\linewidth]{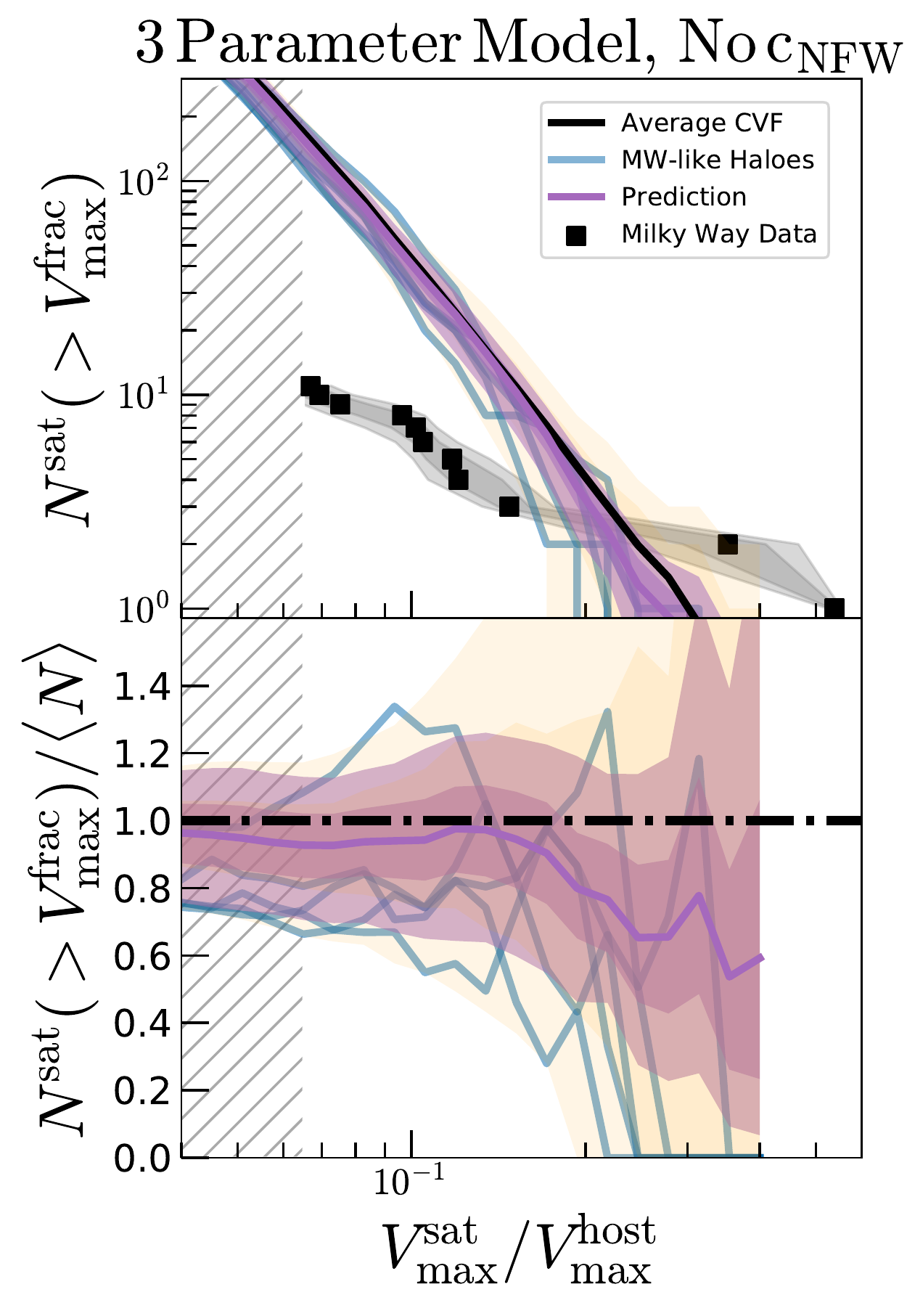}
\caption{As \autoref{fig:CVF3param} but a three-parameter scaling relation which predicts subhalo abundance as a function of spin, shape, and scale factor at the last major merger (we have essentially exchanged $c_{\rm NFW}$ for $a_{\rm LMM}$). The predictions for this model are substantially different from the models that include concentration. This model predicts that the Milky Way's host halo should have up to 17\% fewer to 2\% more ubhaloes than average at the low $V_{\rm max}^{\rm frac}$ end (at 68\% confidence), and 4--35\% fewer subhaloes at the high $V_{\rm max}^{\rm frac}$ end. We emphasize that the blue lines denote the same haloes discussed in \autoref{sub:neighbors} and shown in previous plots (\autoref{fig:triangle}, \autoref{fig:CVF1param},\autoref{fig:CVF3param}, etc). Namely these are the haloes most similar to the Milky Way across all four parameter spaces, which includes concentration.}
\label{fig:CVF3param_noc}
\end{figure}

\autoref{fig:CVF1param}, \autoref{fig:CVF3param}, and \autoref{fig:CVF3param_noc} depict the results of our model fits, evaluated using the properties of the Milky Way host halo determined in \autoref{sub:comparison}. The purple lines show the predicted cumulative velocity functions (i.e., the subhalo abundance for each threshold in velocity fraction considered) for the Milky Way from the one- and three-parameter models, respectively. The dashed grey region indicates the resolution limit of the simulations (which begins to have effects below $V_{\rm max}^{\rm sat}/V_{\rm max}^{\rm host} = V_{\rm max}^{\rm frac} = 0.065$). The solid black line shows the average CVF for the 45 host haloes. In the bottom panels the dashed-dot black line at $N_{\rm sat}(>V_{\rm max}^{\rm frac})/\langle N\rangle = 1$ is shown for comparison to indicate no separation between the predictions and the average CVF depicted by the solid black line (similar to \autoref{fig:CVFs}).

Over-plotted in blue are the cumulative velocity functions for the same 5 nearest neighbours to the Milky Way that were indicated in \autoref{fig:triangle}, \autoref{fig:1param}, and \autoref{fig:3param} for comparison.  The CVF for the Milky Way classical satellites is again depicted by the black points, as discussed in \autoref{section:satellites}.

Uncertainties on our model predictions have been obtained via bootstrap re-sampling. Specifically, we randomly select a set of size $N=45$ of our simulated Milky Way-mass host haloes with replacement (i.e., allowing the same host halo to be selected more than once) and compute the maximum likelihood fit again 10,000 times. For each bootstrap sample, we also draw a new set of Milky Way halo properties from the probability distributions for each defined by their uncertainties (Gaussian for spin, shape, and scale factor at the last major merger, and two half-Gaussians for concentration due to asymmetric errors, see \autoref{Section:maxlikelihood}). We can identify 68\% and 95\% confidence intervals from these bootstrap samples as the regions containing the middle 68\% and 95\% of bootstrap results at a given velocity, respectively. The resulting confidence intervals, which reflect only uncertainties in Milky Way halo properties and the parameters of the subhalo abundance fits, are depicted by the darker and lighter purple regions. 

Additionally, with this bootstrapped sample we can investigate the enlarged range of observed values expected from Poisson scatter about the mean subhalo abundance. For every bin in each bootstrap sample, we randomly generate a value from a Poisson distribution with mean given by the predicted number of subhaloes within that bin ($N_{\rm sub,l}^{\rm pred}(>V_{\rm max, l}^{\rm frac}) - N_{\rm sub,l+1}^{\rm pred}(>V_{\rm max, l}^{\rm frac})$, where $l$ indicates velocity bin number and $N_{\rm sub,l}^{\rm pred}(>V_{\rm max, l}^{\rm frac})$ is the predicted cumulative number of subhaloes down to the minimum velocity of that bin). To generate a CVF for that bootstrap sample we then add together the randomly-generated values cumulatively, starting with the highest velocity bin. The resulting values incorporate both the bootstrap uncertainties and the Poisson scatter in subhalo abundances, with the effects of covariance between velocity bins resulting from our use of cumulative counts properly accounted for.  The $68\%$ and $95\%$ confidence intervals derived from the distribution of bootstrap values with Poisson scatter included are depicted by the darker and lighter orange regions, respectively.

\begin{figure*}
\includegraphics[width=\linewidth]{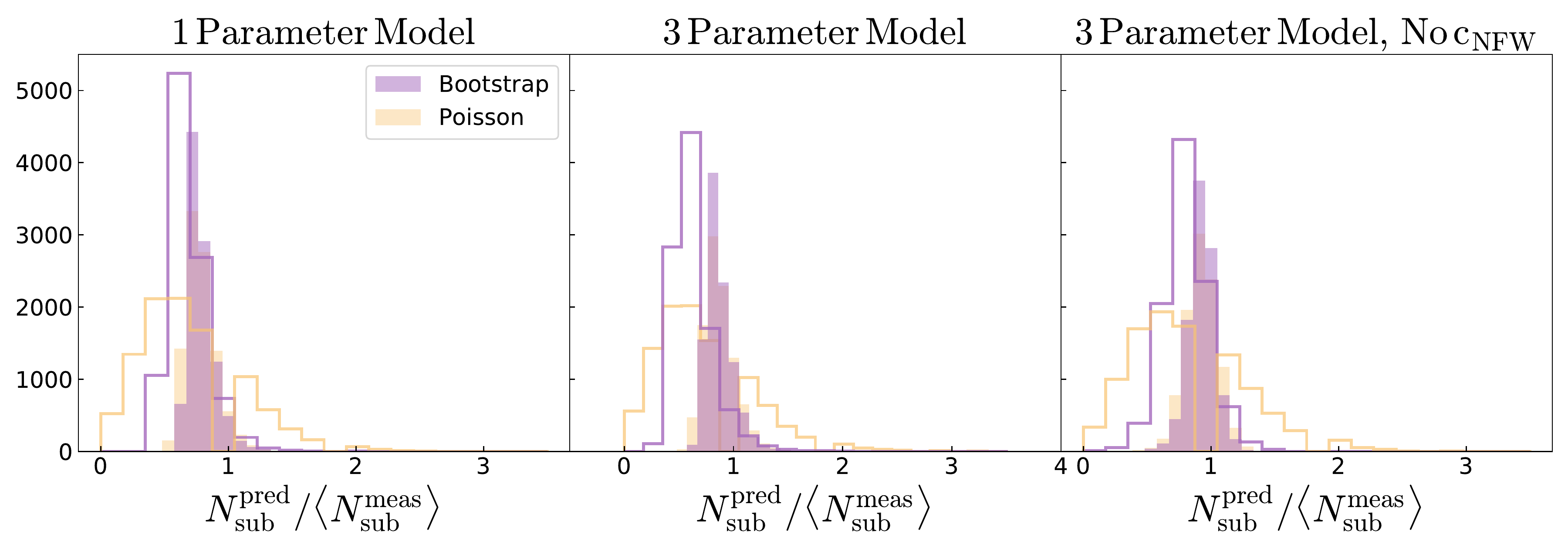}
\caption{Histograms of predicted total cumulative subhalo abundance below the resolution limit normalized by the mean measured subhalo abundance, where errors constituting the purple histograms include uncertainties in fitting power-law models and uncertainties in the Milky Way host halo parameters (based off just the bootstraps), whereas the orange histogram also incorporates uncertainties due to Poisson scatter about the bootstrapped relations. The filled histograms correspond to the low $V_{\rm max}^{\rm frac}$ end just above the resolution limit, and the outlined histograms correspond to the high $V_{\rm max}^{\rm frac}$ end before the subhalo counts are dominated by 0. The incorporation of Poisson scatter leads to a similar spread in predicted subhalo abundance at low $V_{\rm max}^{\rm frac}$ and a much larger spread at hight $V_{\rm max}^{\rm frac}$. The $68\%$ and $95\%$ confidence regions derived from these distributions are provided in \autoref{table:errors}. The one-parameter model predicts slightly fewer subhaloes for the Milky Way than the three-parameter model,which both predict fewer subhaloes for the Milky Way than the model that does not include concentration.}
\label{fig:histerror}
\end{figure*}

The distributions of predicted subhalo abundances for the lowest-velocity bin above the resolution limit ($V_{\rm max}^{\rm frac} > 0.065$) considering only uncertainties in Milky Way parameters and model fits (purple) or including Poisson variation as well (orange) are shown in \autoref{fig:histerror}. In this plot, shaded regions depict distributions just above the resolution limit and outlined histograms correspond to a bin at the high $V_{\rm max}^{\rm frac}$ end before bin counts are dominated by zero. The extents of the 68\% and 95\% confidence intervals corresponding to these histograms are listed in \autoref{table:errors}.

\begin{figure}
\includegraphics[width=1.0\linewidth]{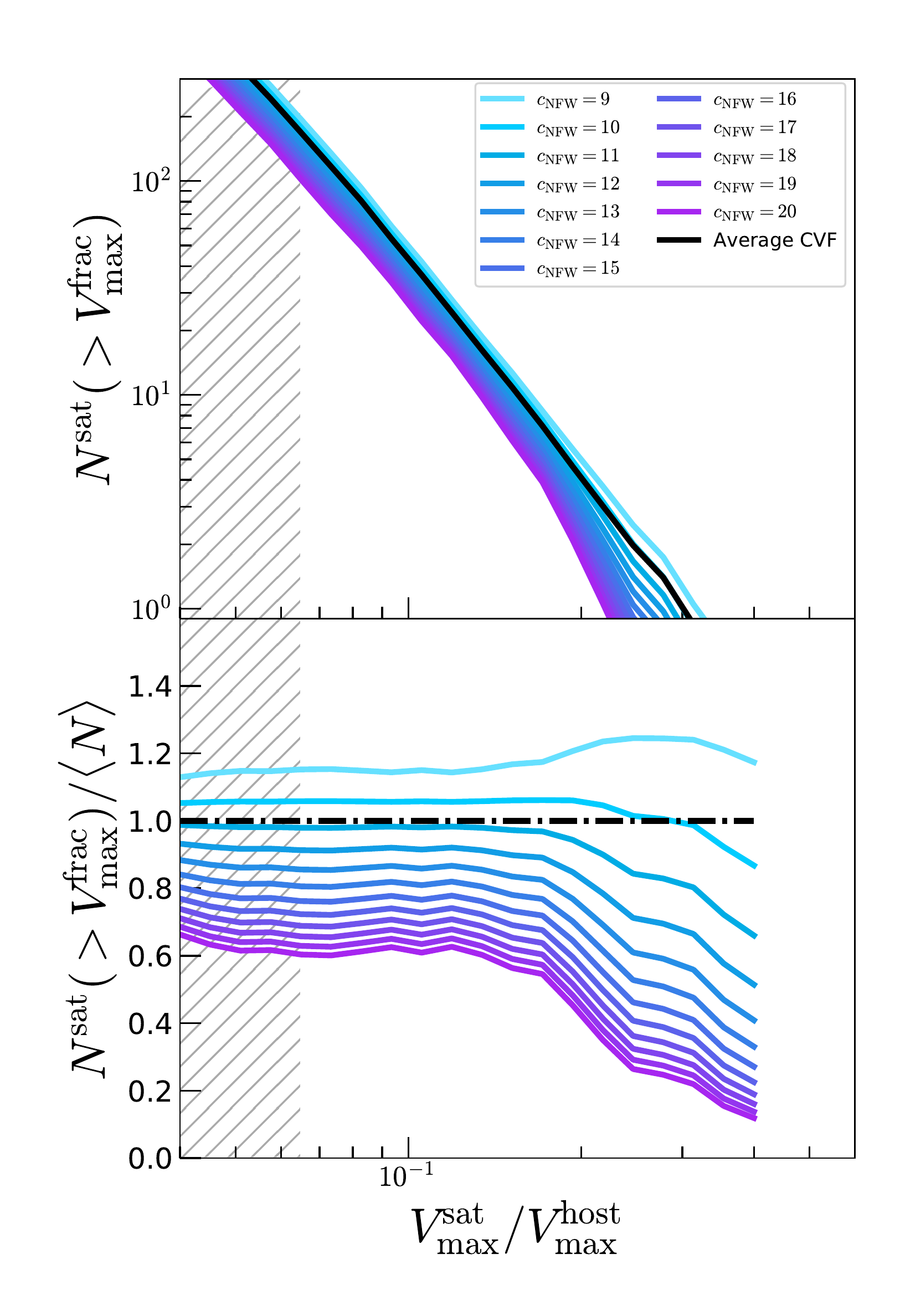}
\caption{A CVF of one parameter models depicted with all possible theoretical values of concentration for haloes at the same mass and morphology as the Milky Way (see \autoref{sub:concentration}). The model is computed in exactly the same way as for \autoref{fig:CVF1param}, and then plotted for various concentration values. haloes of concentration greater than $\sim12$ are consistent with having fewer subhaloes. The high-$V_{\rm max}^{\rm frac}$ tail is more sensitive to the concentration than the overall normalization.}
\label{fig:CVF_allc}
\end{figure}

\begin{figure}
\includegraphics[width=1.0\linewidth]{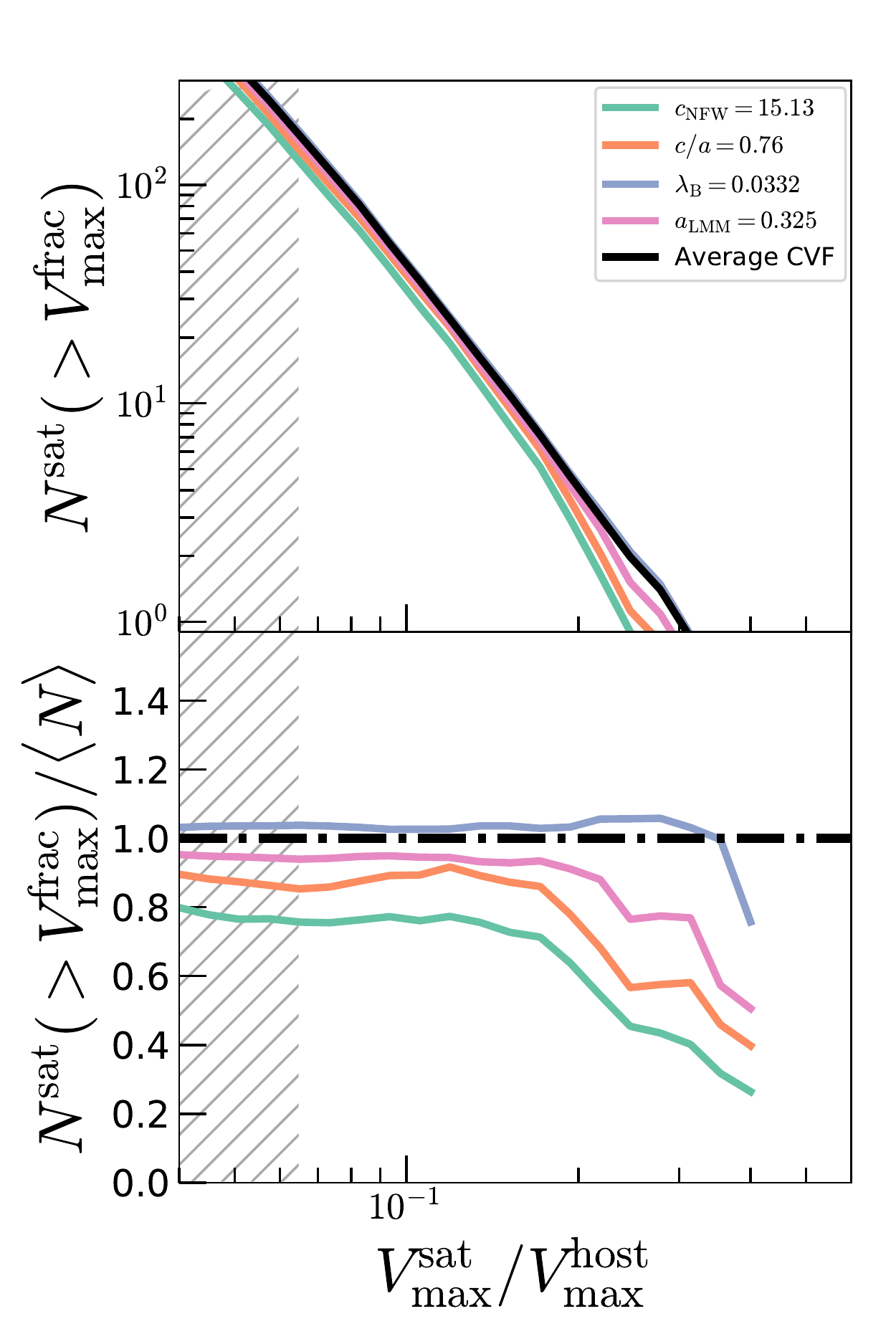}
\caption{A CVF of one parameter models depicted with estimated values of Milky Way host halo parameters (concentration in green, shape in orange, spin in blue, and scale factor at the last major merger in pink), as described in \autoref{Section:milkyway}. It is clear that concentration is the strongest predictor for a far from average subhalo abundance, which is expected given \autoref{fig:CVF_allc}, and the tight correlation between concentration and subhalo abundance seen in \autoref{tab:spearman}. The next strongest deviation comes from shape, followed by scale factor at the last major merger, and lastly spin (which is very close to the average).}
\label{fig:CVF_MW}
\end{figure}

\begin{table}
\centering
\begin{tabular}{|l|l|l|}
\hline
Uncertainty & 68\% & 95\% \\ \hline
One-para. Bootstraps & 0.70--0.88 & 0.64--1.05 \\
Three-para. Bootstraps & 0.76--0.99 & 0.69--1.19\\
Three-para. No $c_{\rm NFW}$ Bootstraps & 0.83--1.02 & 0.71--1.13\\
\hline
One-para. Poisson & 0.68--0.90 & 0.59--1.09 \\
Three-para. Poisson & 0.74--1.01 & 0.64--1.22\\
Three-para. No $c_{\rm NFW}$ Poisson & 0.29--1.05 & 0.62--1.17\\
\hline
\end{tabular}
\caption{Table of $68\%$ and $95\%$  confidence intervals for the total abundance of subhaloes above the resolution limit ($V_{\rm max}^{\rm frac}>0.065$) predicted for the Milky Way (or what we refer to in the text as \textit{low $V_{\rm max}^{\rm frac}$}).  All calculations are done using the power-law models defined by \autoref{equation:power1}. We consider separate confidence intervals for the mean subhalo abundance of a Milky Way-like halo, which include only the uncertainties from subhalo abundance fitting and Milky Way halo parameter uncertainties (labeled as 'Bootstrap' confidence intervals here); and confidence intervals which also include the impact of Poisson scatter about that mean abundance (labeled as  'Poisson'). We provide results for the one-parameter model, the three-parameter model, and three-parameter model without concentration}. All confidence intervals are calculated from the total cumulative predicted subhalo abundance above the resolution limit of $V_{\rm max}^{\rm frac} = 0.065$ divided by the mean total measured subhalo abundance ($N_{\rm sub}^{\rm pred}/\langle N_{\rm sub}^{\rm meas}\rangle$) above this limit. These confidence intervals correspond directly to the shaded regions depicted visually in \autoref{fig:histerror}.
\label{table:errors}
\end{table} 

In both models that include concentration the predicted CVF for the Milky Way lies below the average for the Milky Way-mass dark matter haloes up to $1\sigma$. However, this is not the case for the model that does not include concentration. This means that concentration is a critical component when examining subhalo abundances in relation to host halo properties, which is evident from \autoref{fig:CVF_allc} and \autoref{fig:CVF_MW} (which is also not surprising since, as discussed in \autoref{sub:props_and_abundance} and seen in \autoref{tab:spearman} concentration is by far the parameter most correlated with subhalo abundance). The former depicts one-parameter models based on the total theoretical range of concentration values of a halo of similar mass or morphology to the Milky Way (see \autoref{sub:concentration}. These models are identical to the one-parameter model, but a different concentration is used for plotting (i.e., the $x_{i}$ in \autoref{equation:power1}). The mean concentration of haloes in our simulations is 11.63. A halo with a concentration larger than that mean is consistent what having a noticeably smaller predicted subhalo abundance. Additionally we find that the tail at the high $V_{\rm max}^{\rm frac}$ end is more sensitive to concentration than the overall normalization (concentrations smaller than 9 predict more massive subhaloes than average). The latter depicts one-parameter models for the estimated values of the Milky Way host halo properties as described in \autoref{Section:milkyway}. This shows that concentration is the strongest predictor for a Milky Way-like halo to having fewer subhaloes, followed by shape, scale factor at the last major merger, and spin last.

Based upon the two superior models, we can 
conclude that the Milky Way host halo could have fewer subhaloes than would be typical of a halo of its mass (or should, if we take the mean to be representative). At the low $V_{\rm max}^{\rm frac}$ end, at 68\% confidence not accounting for Poisson scatter we should expect 12--30\% fewer subhaloes than average based on the one-parameter model, or up to 22\% fewer subhaloes based on the three parameter model, when incorporating uncertainties in the subhalo abundance fits and Milky Way parameter values. These percentage ranges are similar, but larger at high $V_{\rm max}$, when Poisson scatter is accounted for. At the high $V_{\rm max}^{\rm frac}$ end the mean prediction is even smaller compared to the average (19--52\% fewer subhaloes), but uncertainties in fit parameters and fractional Poisson scatter are greater, making the presence of satellites as large as the Magellanic Clouds rare but not extraordinary. Additionally, despite the three-parameter model without concentration being much less robust (see \autoref{table:AICBIC}), it still predicts as many as $17\%$ fewer subhalo at low $V_{\rm max}^{\rm frac}$ and up to $35\%$ fewer subhaloes at high $V_{\rm max}^{\rm frac}$ at $1 \sigma$.

The differences between the predicted cumulative velocity functions of the two concentration based models are relatively small. This also demonstrates that a simple concentration-based model is adequate for describing subhalo abundances in Milky Way-like haloes to first order, consistent with the results from \citet{mao2015} and \autoref{fig:CVF_MW}. This is no surprise, given $c_{\rm NFW}$'s close relationship to the other parameters. For example, when we repeat the same analysis with a one-parameter model using the next-most correlated property in \autoref{tab:spearman}, $a_{\rm LMM}$, only $0-20\%$ fewer subhaloes than average are predicted at low $V_{\rm max}^{\rm frac}$, while the results are unstable at high $V_{\rm max}^{\rm frac}$. $c_{\rm NFW}$ is significantly more tightly correlated with $N_{\rm sub}$ than the other host halo properties considered.

As discussed in \autoref{sub:models}, the three-parameter model predicts subhalo abundances for the simulated haloes with smaller scatter than the one-parameter model.  As a result, at first glance one would expect this model to also yield more compact confidence intervals for the abundance of subhaloes around the Milky Way, but in \autoref{table:errors} the opposite holds true (e.g., the 95\% confidence interval including all sources of scatter spans 41\% in the one-parameter model versus 50\% in the three-parameter model with concentration). This can be explained at least in part by the uncertainties in Milky Way halo properties beyond concentration, the effects of which will alter the confidence intervals for the three-parameter model, but not for the one-parameter case. Specifically, the three-parameter will be a better model in the future when there are better constraints on the properties of the Milky Way. Otherwise the errors on the parameters propagate into the errors of the subhalo abundance estimate.

\autoref{fig:CVF1param}, \autoref{fig:CVF3param}, and \autoref{fig:CVF3param_noc} make clear that at $V_{\rm max}^{\rm frac}$ above $\sim$0.2 the typical halo whose properties match the Milky Way's will have even fewer subhaloes compared to the average - a deficiency in excess of 60\%. This makes it somewhat more surprising that the MW should have any relatively large satellites such as the Large and Small Magellanic Clouds (LMC and SMC) or Sagittarius (the SMC and Sagittarius correspond to the two highest-$V_{\rm max}^{\rm sat}$ points in each figure; the LMC is off the plot range with $V_{\rm max} = 91.7$ km\,s$^{-1}$), as more massive satellites are rarer in the haloes most like the one which hosts our Galaxy.  

With these results we can define fits for the parameter exponents, $\alpha_{i}$, and the scale $k$ as functions of $V_{\rm max}^{\rm frac}$ for both the one-parameter and three-parameter models which include concentration, as those are the superior models. These functions can be used in order to determine a CVF for any set of host haloes. The functions and process is described in \autoref{sub:generalfunc}. 

\subsection{Halo Properties and Subhalo Scaling Relations}
\label{Section:tbtf}

As briefly mentioned previously, the issue of 'too-big-to-fail' (TBTF) refers to the overabundance of specifically massive and dense subhaloes predicted from CDM simulations in comparison to the number of luminous satellites that the Milky Way has been observed to host \citep[etc.]{b-k2011,b-k2012,tollerud2014}. The TBTF has been formulated in many different ways; the most well known is through $N$-body simulations that show Milky Way-mass hosts should host 6--10 subhaloes with a potential well depth of $V_{\rm max} > 25$ km\,s$^{-1}$ \citep[e.g.,][]{madau2008,b-k2011,jiang2015}. This is not consistent with observations around the Milky Way; only the LMC and SMC fit into this criteria. Another formulation of TBTF is in terms of density. The central densities of Milky Way satellites that have been inferred from kinematics are too low compared to the central densities of their dark matter subhalo counterparts \citep{b-k2011,purcell2012,jiang2015}.

In quantitative terms, the density formulation of the TBTF is the statement that the classical dwarfs imply a larger radius at which the maximum velocity is reached ($R_{\rm max}$) for a given maximum velocity ($V_{\rm max}$) than is typical of CDM subhaloes, or that the subhaloes are denser then their kinematics suggest. We use $R_{\rm max}$ and $V_{\rm max}$ because NFW profiles consist of two parameters, and these are easily related to observed stellar kinematics. Subhaloes fall on a narrow line in $R_{\rm max}$--$V_{\rm max}$ space, seen in \citet{zentner2003} and subsequent papers. If the structural relation of $R_{\rm max}$ and $V_{\rm max}$ varies systematically as a function of host halo property (as is the case in our CVFs), this could hold an interesting implication for TBTF --- an implied relationship between subhalo density and host halo property.

We have used our model for satellite abundances in Milky Way-like haloes to investigate whether the TBTF issue can be related to host halo properties. Specifically, we have measured the relationship between the maximum circular velocity of subhaloes and their maximum radius, dividing haloes up into quartiles based on host halo properties as above. However, the results were inconclusive: i.e., any separation in the $R_{\rm max}$-$V_{\rm max}$ plane for samples divided into dark matter host halo property quartiles is relatively weak. We find no statistically significant separation when dividing host haloes into quartiles according to their halo concentration, spin, shape, or major merger scale. Any unusualness in the properties of the Milky Way halo cannot be used to explain structural differences between the observed satellite properties and the expected characteristics of subhaloes; rather, we find the best-fit scaling relations to be essentially independent of halo properties. Improvements to this analysis would require a much larger sample of Milky Way-mass haloes to be re-simulated at high resolution.

\section{Conclusion}
\label{Section:conclusion}

In this paper, we have utilized Milky Way-mass zoom-in simulations to investigate the sources of the scatter in subhalo abundances at fixed mass (see \autoref{fig:CVFs}). We particularly focus on predicting subhalo abundance conditioned on properties of a host halo. Recent studies of the Milky Way have revealed that the Milky Way has an unusually small disk \citep{Licquia2015,bland2016,licquia2016}, which in standard galaxy formation theory would be related to unusual host halo properties; we have sought to determine if the particular properties of our Galaxy's halo would also cause its expected satellite galaxy abundance to be unusual. The aspects of the halo we have investigated are its concentration ($c_{\rm NFW}$), spin ($\lambda_{\rm B}$), shape ($c/a$), and scale factor at last major merger ($a_{\rm LMM}$), as discussed in \autoref{Section:milkyway}.

First, we conclude that based on current estimates of its properties the Milky Way's host dark matter halo indeed lies at an extrema compared to haloes of its mass from N-body simulations (see \autoref{fig:triangle}). In particular, the Milky Way lies away from the median in the projections across the full parameter space (higher-than average $c_{\rm NFW}$, lower than average $\lambda_{\rm B}$, more spherical than average $c/a$, and a very small $a_{\rm LMM}$). Next, from our N-body simulations we have determined that the host halo properties considered are significantly correlated with subhalo abundances for Milky Way-mass dark matter haloes (\autoref{tab:spearman}); as a result, they can be used to predict the cumulative velocity function of a given halo (\autoref{fig:CVFs}). haloes with lower-than average concentration host a greater number of subhaloes than haloes with higher-than average concentrations. Similarly, lower-than average spin haloes host fewer subhaloes than higher-than average spin haloes, lower-than average (less spherical) shaped haloes host fewer subhaloes than higher-than average shaped haloes, and earlier forming haloes host fewer subhaloes than later forming haloes. In concordance with estimates for the Milky Way, it should be expected that the Milky Way should host fewer subhaloes.

Using the results from the simulations, we have built two sets of scaling-relation models that predict subhalo abundance above a given threshold velocity based upon the properties of a dark matter halo. In the first model, we predict the subhalo abundance based on a single parameter (at fixed halo mass), namely concentration ($c_{\rm NFW}$). Our second model was a three-parameter model that conditioned subhalo abundance on $c_{\rm NFW}$, spin ($\lambda_{\rm B}$), and host halo shape ($c/a$). We also compare to a three-parameter model that does not include concentration, due to the current limited understanding of halo contraction making the Milky Way concentration a rough estimate. We then evaluate these models with the estimated properties of the Milky Way's host dark matter halo to predict subhalo abundances for our Galaxy. 

The conclusion of this analysis is that we should expect a host halo similar to the Milky Way's to possess fewer subhaloes than the average halo of its mass. However, the error on actual measurements of the Milky Way dark matter halo also make the range of predicted subhaloes consistent with no effect. We will focus our discussion on the impact of the likely scenario that a halo like the Milky Way has fewer subhaloes when utilizing host halo parameters beyond mass. The central predicted values of our best models are well below the mean, which implies a significant probability of a subhalo count deficit. 
This result is summarized in Figures ~\ref{fig:CVF1param} and ~\ref{fig:CVF3param}. Both classes of model yield the same basic result: the Milky Way is predicted to have 1--30\% fewer subhaloes at low circular velocities ($V_{\rm max}^{\rm frac}$) and 19--52\% fewer at high circular velocities than a typical halo of its mass, at $68\%$ confidence when considering only model fitting and Milky Way parameter uncertainties. The decrement with respect to the average cumulative velocity function of dark matter haloes is itself a function of $V_{\rm max}$ (i.e., a function of subhalo mass). The effect is much larger than estimated uncertainties in the fitting and in propagated Milky Way parameters. Current observations have detected approximately 26 satellites around the Milky Way with $V_{\rm max}^{\rm sat} > 10$ km/s (this does not include completeness correction or any other corrections). The mean prediction of the one- and three-parameter models estimate 127--135 total subhaloes around the Milky Way, of which some fraction should host observable satellites. At 1$\sigma$ below the mean our models predict as few as 118--129 subhaloes around the Milky Way. In comparison the mean subhalo number for the Milky Way mass host haloes in our zoom-ins is 169 total subhaloes with $V_{\rm max}^{\rm sat} > 10$ km/s. The similar results from both models indicates that the dominant effect is the relationship between halo concentration, $c_{\rm NFW}$, and satellite abundance, $N^{\rm sat}$, as exemplified by \autoref{fig:CVF3param_noc},\autoref{fig:CVF_allc}, and \autoref{fig:CVF_MW}. It appears that a $c_{\rm NFW}$-based model is generally adequate for predicting subhalo abundance in the mass range of the Milky Way's dark matter halo, though more complicated models can yield smaller errors.

Additionally, we have found that variations in host halo properties do not have a statistically significant impact on the structure of dark matter subhaloes themselves (at least as assessed using  properties connected to TBTF within our sample; see \autoref{Section:tbtf}). Only the subhalo numbers (and hence the MSP) have been impacted by taking host halo properties into account. A set of new halo resimulations at a  resolution $8\times$ higher than those used in this work are now under way and may enable improved investigation of the TBTF problem.

The results described above suggest that a non-negligible fraction of the 'missing satellites' problems is a result of the unusual formation history of the Milky Way. The halo of the Milky Way formed early with very few recent major mergers, which resulted in a more spherically-shaped halo. This also would be expected to lead to a  more centrally-concentrated dark matter halo --- consistent with the estimates shown in \autoref{fig:triangle} as well as results from, e.g., \citet{wechsler2002,zhao2003} and \citet{ludlow2016}. 
This lack of major mergers should also lead to a relatively small angular momentum of the Milky Way halo. This is consistent with previous results showing that at all masses, haloes with lower spin tend to be in less-dense regions and less strongly clustered \citep{gao2007,faltenbacher2010,villarreal2017}, such that many reside in environments resembling our Local Group. All of these halo characteristics correlate with having a smaller number of satellites.  The results of our models are all consistent with scenarios where the Milky Way's low satellite abundance compared to simple $\Lambda$CDM predictions may in part be related to its quiet accretion history as was speculated in \citet{licquia2016}.

However, the low subhalo abundance we predict for the Milky Way dark matter halo based on its properties is not on its own sufficient to explain the missing satellites issue. Other factors, such as baryonic physics, must still play a role. In \autoref{Section:intro} we listed several of the numerous solutions to the MSP. We will discuss how our results tie in with those solutions below.

\begin{enumerate}
\item \textbf{Baryonic effects:} Although our work has shown that we should expect there to be fewer subhaloes than previously anticipated for the Milky Way, the observations and predictions still do not match. At the low velocity end we predict 5$\times$ as many subhaloes as have been detected to date.  Baryonic physics that causes any satellites in these subhaloes to be difficult to detect could address this problem; the strength of baryonic effects required would be smaller than previously estimated, however. Our results suggest that the Milky Way begins with a state of up to $\sim 30\%$ fewer small subhaloes and up to $\sim 50\%$ fewer larger subhaloes than average. Baryonic effects do not need to be as efficient as proposed and can use insight from host halo parameters to better tune models.

\citet{zolotov2012} and \citet{brooks2013} suggest that a combination of supernova feedback and enhanced tidal disruption caused by the presence of a baryonic disk can resolve the missing satellites problem. However, simulations which recover a Milky Way-like population of satellites in a \textit{typical} galaxy of Milky Way mass must be somewhat mis-tuned. That is, either feedback or tidal effects from the baryonic disk must be weaker than was assumed in the simulations of \citet{brooks2013} in order to get a set of satellites like those that surround our Galaxy when starting out with fewer subhaloes than are typical.   

Concentration is an indicator of formation time that may contain more information about the global formation of a halo than simply the time of the last major merger. In host haloes that accrete their substructure earlier, tidal stripping will have longer to operate, depleting subhalo abundances. This implies that the subhaloes within our Galaxy's halo were likely largely accreted relatively early compared to those surrounding other galaxies (barring those associated with the LMC and SMC which may be on their first infall \citep{besla2007,besla2010}) within the mass range of the Milky Way's halo. This may cause tidal stripping to have stronger effects in the Milky Way system than is typical for a galaxy of its mass. 

\item \textbf{Non-cold dark matter and exotic physics:} Similar arguments apply to more exotic physics as to baryonic effects. If, for example, a non-CDM dark matter model and/or some modification to inflation were invoked to alleviate the MSP and TBTF issue, those modifications would need to be weaker than previously assumed, given the smaller Milky Way subhalo abundance that we predict.
\end{enumerate}

To summarize, when exploring potential factors affecting the missing satellites problem and tuning models to match the Milky Way's satellite population, it is important to ensure that those models predict that a galaxy will have a Milky Way-like satellite population, not for an average halo of Milky Way host mass, but rather for one which has Milky Way-like properties across the board.  
 
We note that uncertainties on the mass of the Milky Way's dark matter halo are sufficient that it may be as small as one half of the value assumed when selecting haloes for resimulation in \citet{mao2015}. Since the number of subhaloes is proportional to halo mass in the parent simulations from which those haloes were drawn, we would correspondingly expect a 50\% smaller subhalo population for a typical galaxy of Milky Way halo mass than what the simulation results give \citep{wang2011}.  In combination with the differences between the expected population of Milky Way subhaloes and the population in a more typical galaxy implied by our model fits, this would mean in net that a 3--4$\times$ reduction in the number of subhaloes of the Milky Way halo (and hence the strength of the missing satellites problem) compared to what was previously assumed is entirely possible.  

One outstanding anomaly is that, although the Milky Way overall has fewer satellites than might be expected for the typical halo of its mass, it actually has more of the most massive satellites than average.  Previous numerical and observational studies have estimated that there is a 2.5--11\% chance that a halo of the mass of the Milky Way hosts two subhaloes as large and luminous as the Magellanic Clouds (e.g., \citet{busha2011}). Taking our Galaxy's halo properties into account only increases this contrast.  \autoref{fig:CVF1param}, \autoref{fig:CVF3param} and \autoref{fig:CVF3param_noc} show that at $V_{\rm max}^{\rm frac}$ above 0.2 the Milky Way subhalo abundance is expected to be further below the average for a halo of its mass than at lower velocities (with a decrement of up to 50\% of the standard prediction, considering only model fitting and Milky Way parameter uncertainties).  This makes the 'too-big-to-fail' problem even more of a puzzle. The Milky Way’s host halo properties predict in our model that these large companions are very unlikely in concordance with TBTF, yet the Milky Way hosts two very massive companions. However, we note that the large Poisson scatter in this regime tends to dwarf the suppresion of subhalo abundances. It would be interesting to follow up this work with an enlarged suite of re-simulated haloes that is more well-suited to addressing the abundances of these large, rare subhaloes.

There are several important caveats to our results which are important to keep in mind when interpreting this work. First, existing constraints on the Milky Way's halo properties are in many cases only rough estimates; this can propagate through into predictions of satellite populations. We have discussed the issue of concentration in detail in \autoref{sub:concentration} and \autoref{Section:Adiabatic Contraction}, where adiabatic contraction results in higher concentrated and less NFW-like haloes which is an issue that has not been explored in detail in the literature. Additionally one might worry particularly about estimates of our Galaxy's spin parameter, as there is significant doubt that the connection between disks and haloes is as tight as implied by the \citet{mmw1998a} model. Simple models of galaxy formation assume that angular momentum conservation during collapse leads spiral galaxies to have specific angular momenta comparable to the haloes they reside in \citep{fall1980,mmw1998a,bullock2001}. However, simulation work by \citet{jiang2018} shows that the halo spin parameter is not significantly correlated with galaxy size. 
Nonetheless, we remind the reader that it is concentration, and not spin, that drives the bulk of the reduction in subhalo abundance at fixed halo mass; changing the estimated spin parameter by a factor of two has a $\sim 5\%$ effect on the predicted subhalo abundance at low $V_{\rm max}^{\rm frac}$ and no noticeable effect at high $V_{\rm max}^{\rm frac}$.

In this paper, we have focused entirely on the problem of differences between the subhalo abundance in Milky Way-like haloes versus the typical halo of the same mass.  However, in exploring this problem we have investigated two related issues which we will focus on in upcoming papers.  First, we have investigated the influence subhaloes have on measurements of dark matter halo properties in simulations. This will provide us with a better understanding of the quantities used as input for semi-analytic models, as well as potential differences between estimates of halo properties for real galaxies versus measurements of those properties in simulations. Second, we have explored the relationship between subhalo abundances and the galaxy-halo connection. 
An interesting extension of this work which we have not pursued to date would be a comparison of subhalo abundances with the observed population of satellites around M\,31, or with satellite populations around Milky Way analog galaxies. Disk scale length (compared to expectations from the mass-size relation of galaxies) could potentially be used to separate high-spin from low-spin haloes of comparable mass, enabling an empirical test of whether satellite abundance correlates with halo properties beyond mass.  Through such studies, data from current and upcoming surveys \citep[e.g., the SAGA Survey;][]{geha2017} have the potential to provide us with more insight into the Missing Satellites and Too Big to Fail problems. 

\section*{Acknowledgements}
This research made use of computational resources at SLAC National Accelerator Laboratory, a U.S.\ Department of Energy Office; YYM thanks the SLAC computational team for their support. 
ARZ is supported by the Pittsburgh Particle physics Astrophysics and Cosmology Center (PITT PACC) at the University of Pittsburgh and by the U.S. National Science Foundation (NSF) through grants 
NSF AST 1517563 and NSF AST 1516266. YYM is supported by the Samuel P.\ Langley PITT PACC Postdoctoral Fellowship. 

This research made use of Python, along with many community-developed or maintained software packages, including
IPython \citep{ipython},
Jupyter (\http{jupyter.org}),
Matplotlib \citep{matplotlib},
NumPy \citep{numpy},
Pandas \citep{pandas},
and SciPy \citep{scipy}.
This research made use of NASA's Astrophysics Data System for bibliographic information.

\bibliographystyle{mnras}
\bibliography{refs,software}

\clearpage

\appendix
\section{Concentration and Adiabatic Contraction}
\label{Section:Adiabatic Contraction}

In order to compare the estimated concentration of the Milky Way to our dark matter only data, we must explore how adiabatic contraction of the halo effects the concentration measurement. We do this by employing CONTRA, a publicly available code that calculates the contraction of a dark matter halo as a result of a central population of baryons \citep{gnedin2004}. This code assumes a spherically symmetric distribution of matter for the dark matter halo (which we know is not entirely accurate), and that the velocity distribution is isotropic (which we know is also unlikely to be the case - the velocity distribution varies throughout the halo). However, it is the most simple model to begin with.

We explore results based upon the updated model for halo contraction from \citet{gnedin2004}. The Blumenthal \citep{blumenthal1986} model is the original standard model that has since been updated. The Blumenthal model treats a halo as spherically symmetric, which undergoes homologous contraction - spherical shells that contact in radius but do not cross each other, with particle orbits that are circular, and angular momentum is conserved. The Gnedin model adds in a correction for gas dissipation, which better accounts for effect from mergers and feedback from star formation. It essentially allows for eccentricity in particle orbits that better reflects a complex formation scenario. 

The parameters used for input into CONTRA are those determined for the Milky Way from \citet{bland2016}, were the baryonic fraction $f_{b} = 0.07 \pm 0.001$, the baryon scale length $R_{b} = 0.00957$ incorporating the scale length of the disk from \citet{licquia2016}, and we assume no velocity anisotropy. We calculate an NFW fit to circular velocity where contracted concentration is a free parameter (we also studied a model allowing mass to also be a free parameter which yielded similar results):
\begin{equation}
    V(c_{\rm NFW}) = V_{\rm vir}\sqrt{\frac{1}{x}\frac{\ln{(1+c_{\rm NFW}x)}-\frac{c_{\rm NFW}x}{(1+c_{\rm NFW}x)}}{\ln{(1+c_{\rm NFW})}-\frac{c_{\rm NFW}}{1+c_{\rm NFW}}}}
\label{eq:vnfw}
\end{equation}
where $x = \frac{r}{R_{\rm vir}}$ with r as the contracted positions of the dark matter particles from Contra, $V_{\rm vir} = \sqrt{\frac{GM_{\rm vir}}{R_{\rm vir}}}$, $M_{\rm vir} = 1.3\times10^{12} \msun$ for the Milky Way from \citet{bland2016}, $R{\rm vir} = 282$ kpc, and $G$ is Newton's gravitational constant. Data output from CONTRA is log-spaced in 80 bins, so all of our analysis has the same structure, and the errors are linear.
In order for us to avoid fitting to a more complex model (which would constitute its own paper), we fit in the outer region of the halo ($r >20$ kpc) in rotation curve space in order to mitigate effects from the bulge and the disk. 

\begin{figure}
    \includegraphics[width=1.0\linewidth]{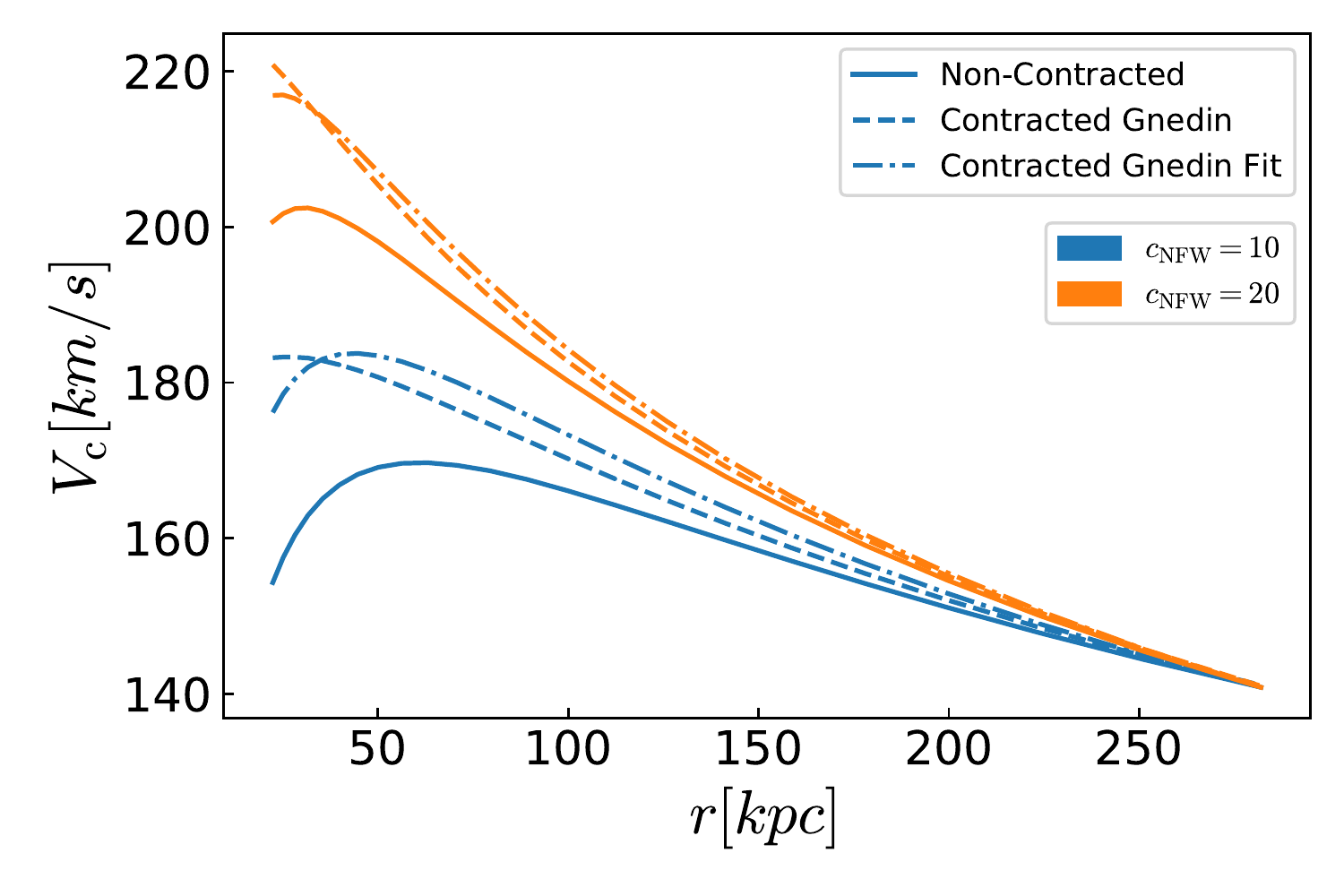}
    \caption{Rotation curves based on the Gnedin contraction model for two initial NFW concentration values of 10 and 20, depicted by blue and orange respectively. These concentration values fully span the theoretical range for haloes of the same mass as the Milky Way. In both cases the solid lines depict what the rotation curve looks like without any contraction, which perfectly matches to the fit of \autoref{eq:vnfw} by design. The over-plotted dashed lines represent the contracted results output from CONTRA, and the dashed-dot lines are the fit to \autoref{eq:vnfw}. It is evident that the fit is not a good depiction of the CONTRA data.}
    \label{fig:rotation_curve}
\end{figure}

This fit was calculated using $\tt scipy.optimize.curve\_fit$ from \citet{scipy}, which is a form of non-linear least squares fitting. The results are presented for two concentration values for the Gnedin model in \autoref{fig:rotation_curve}. The blue curves correspond to an initial (pre-contraction) concentration of 10, and the orange curves correspond to an initial concentration of 20. This spans the approximate theoretical range of NFW concentrations for the Milky Way. In both cases the solid lines depict what the rotation curve looks like without any contraction (\autoref{eq:vnfw}). The over-plotted dashed lines represent the contracted results output from CONTRA, and the dashed-dot lines are the fit to \autoref{eq:vnfw}. It is evident that the fit is not a good depiction of the CONTRA data. This implies that for a fit to contracted data for a halo that most resembles the Milky Way, NFW is not the best model even in the outer regions of the halo where the fit matches the form of the non-contracted model better than the contracted data.

\begin{figure}
    \includegraphics[width=1.0\linewidth]{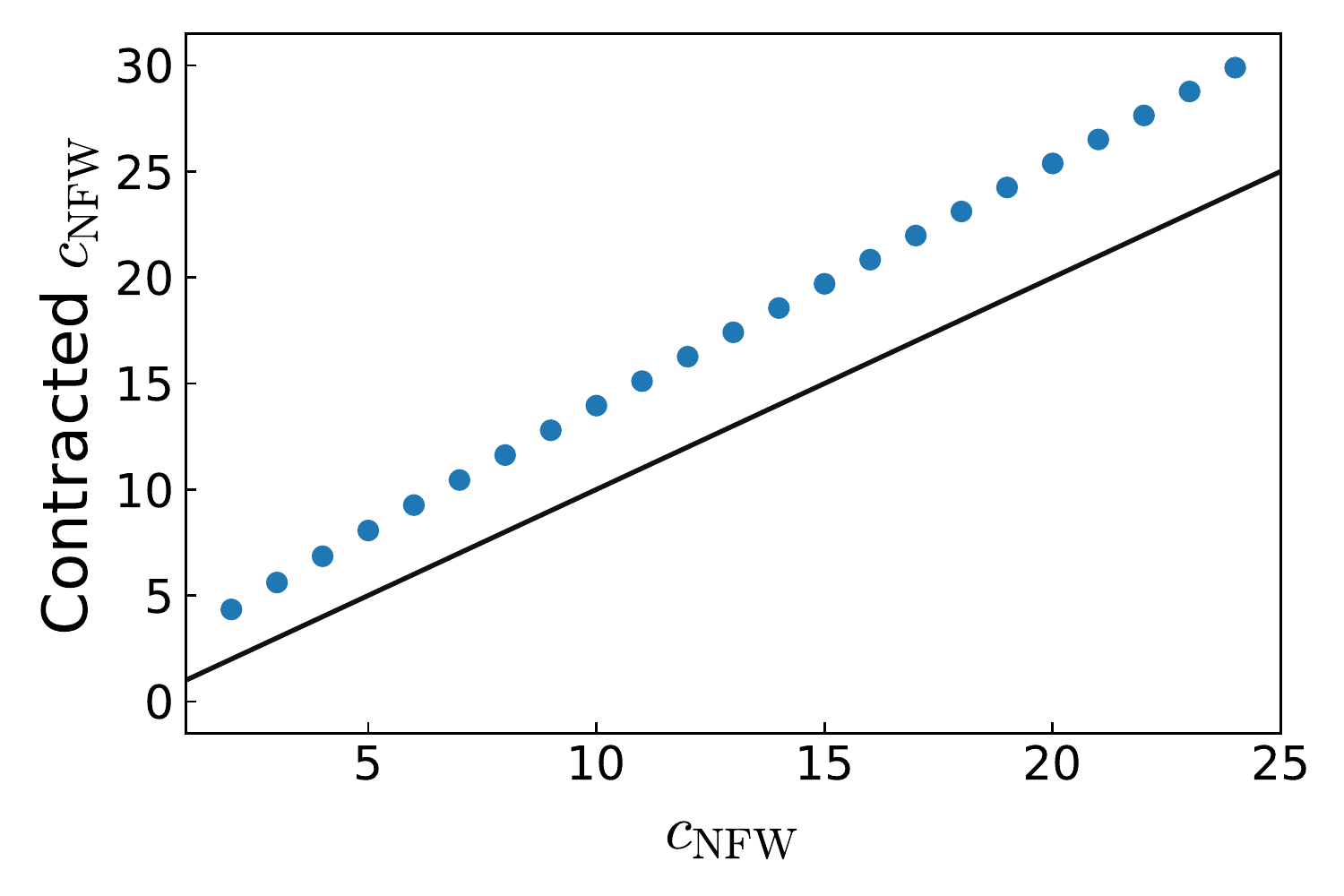}
    \caption{This plot presents the contracted concentration as a result from our fit as a function of the initial NFW concentration values input into CONTRA, denoted by blue points. For comparison the black line shows the 1:1 ratio which would indicate no change in concentration. The differences are relatively small, where, for example, a halo of NFW concentration of 10 is expected to have a concentration of 13.95 after contraction.}
    \label{fig:concentration_scatter}
\end{figure}

\autoref{fig:concentration_scatter} shows the concentrations as a result of the contraction plotted as a function of their initial values. The differences are small but notable. haloes starting with a higher concentration experience slightly more contraction, which is anticipated. Although we reproduce the expected near 1:1 variance in concentration, the fits themselves are poor.

Despite the inaccuracies in the fits to the adiabatically-contracted profiles, we can take the resultant concentration from the fit in order to calculate NFW concentration as a function of contracted concentration. This is done using the \citet{scipy} 1D interpolator. There are a range of estimated values for the Milky Way concentration using various dynamical tracers (a majority being blue horizontal branch stars) and various models. We incorporate results from \citet{battaglia2005,cantena2010,deason2012,nesti2013,kafle2014} and \citet{zhai2018} in order to calculate the best estimate for the Milky Way's concentration post-contraction with a median and median interval of $19.85^{+3.5}_{-5.085}$. We re-map this estimate and errors into non-contracted space, yielding a best result for the Milky Way of $c_{\rm NFW}^{\rm MW} = 15^{+1.35}_{-2.58}$.

 It is erroneous to say that a contracted halo similar to the Milky Way has a concentration in the same sense that an NFW halo does, for the case of a simple model. One reason for the bad fitting could be the level that the Milky Way itself deviates from an NFW profile - an NFW profile is the average for haloes and only good down to $\sim 10-20\%$. Another is the fact that there is velocity anisotropies in the halo. These are issue that needs to be explored further in the field, but is beyond the context of this paper.

\section{Numerical and Mathematical Techniques}
\label{Section:Numerical Techniques}

In the appendices which follow, we discuss details of the numerical and mathematical techniques used in this work. First, in \autoref{sub:regressions} we describe our  exploratory linear regression modelling using a variety of combinations of host halo properties as predictors for subhalo abundance. Next, in \autoref{sub:neighbors}, we discuss the methods used to select the five simulated haloes with properties nearest to those of the Milky Way, which are indicated as blue points or lines in Figs.~\ref{fig:triangle}, \ref{fig:CVF1param}, \ref{fig:CVF3param}, \ref{fig:1param}, \ref{fig:3param}. In \autoref{sub:models}, we discuss the selection of the two models used in our final analyses based upon their information criteria. In \autoref{Section:maxlikelihood}, we discuss in detail the Poisson maximum likelihood methods used to fit power-law scaling models for the subhalo abundance. Lastly in \autoref{sub:generalfunc} we provide a general fitting function for satellite abundance. 

\subsection{Regression Modeling}
\label{sub:regressions}

Our exploration of models for the dependence of subhalo abundance on halo properties starts with simple linear regression modeling. We begin by developing models for the total abundance of subhaloes per host halo, above the resolution limit described in \autoref{sub:simulations} equal to $V_{\rm max}^{\rm frac} = 0.065$. In that regime, there are $>100$ subhaloes per host halo, so Poisson errors in the abundance of subhaloes can be approximated well by a Gaussian distribution.  As a result, if we restrict ourselves to purely linear models for simplicity, ordinary least squares (OLS) regression provides an appropriate analysis technique for our data.  

Specifically, we regress the total number of subhaloes each host has (our dependent variable) against the matrix of host halo properties considered in this paper (constituting the independent variables in this problem). The host halo property distributions are more uniformly distributed in linear space (in the cases of concentration and shape) or log space (for spin, last major merger scale, and mass), so before we take any further steps the log of $\lambda_{\rm B}$ and $a_{\rm LMM}$ are taken. Additionally, before the regression is performed (but after log transformations are applied), each set of properties is normalized to have mean zero and variance one, enabling coefficients of different properties to be directly compared to each other. We utilize the \citet{scikit-learn} $\tt sklearn.linear\_model.LinearRegression$ class for the OLS regression. We can compare the least-squares predictions to the actual number of subhaloes in each halo to test to what degree a given model explains the overall subhalo abundance. 

\subsection{Nearest Neighbors}
\label{sub:neighbors}
Throughout this work we highlight the simulated host haloes that are most similar to the Milky Way in their properties. Specifically, the blue points or curves in our figures correspond to the five nearest neighbors to the Milky Way properties in the multidimensional parameter space of $\lambda_{\rm B}$, $c_{\rm NFW}$, $c/a$, and $a_{\rm LMM}$ (the simulated haloes are already selected to have approximately the same mass). 

We identify these neighbors incorporating the results of the regression fits described in  \autoref{sub:regressions}. Explicitly, we begin by transforming the estimates of Milky Way host halo properties described in \autoref{Section:milkyway} to the same scale as the halo properties used in regression by subtracting off the mean and dividing by the standard deviation of the values of that property amongst the simulated host haloes.  We then multiply each host halo property (and the corresponding Milky Way values) by the regression coefficients for that parameter from an OLS linear regression.  We can then define a 'distance' by the square root of the sum of the squares of the differences of each of these re-normalized and re-weighted properties.  This distance will be smallest for those haloes which most closely resemble the Milky Way \textit{in the properties which most strongly determine subhalo abundances}. 

For convenience we determine the nearest neighbors using the $\tt scipy.spatial.KDTree()$ function from \citet{scikit-learn}.  This function allows neighbors -- that is, those haloes with smallest distance from a given point in parameter space, using the metric defined above -- to be identified using a multidimensional binary tree, also known as a \textit{k}-d tree.

\subsection{Model Selection}
\label{sub:models}
In \autoref{sub:mw_model} we compare predictions for subhalo abundances from two models of differing complexity that include concentration. These were selected out of a set of more than twenty different models incorporating various linear combinations of host halo properties and their products, with up to five parameters per model in total. The parameters for each model were determined using simple linear OLS regression as discussed in \autoref{sub:regressions}. We also compare to an inferior model that does not include concentration because of the current difficult nature of measuring concentration for the Milky Way. 

The Akaike Information Criterion (AIC) and the Bayesian Information Criterion (BIC) are two statistics commonly used  for informing model selection \citep{akaike1974,atkinson1981}. Adding additional parameters will always tend to increase likelihood values (or decrease chi-squared) when fitting models to a given dataset, even if those parameters do not truly have intrinsic explanatory power.  The information criteria  were introduced in order to penalize goodness-of-fit measures based upon the number of free parameters in a model in order to mitigate this issue. We have calculated both quantities for all models investigated to determine which ones provide the best fits to the data given their level of complexity; the two models we focus on in this paper had the lowest AIC and BIC values of any models considered that lack degeneracies when given in power-law form. The information criteria can be defined as
\begin{align}
\text{AIC} &\equiv 2{k} - 2\ln(\hat{L}); \\
\text{BIC} &\equiv \ln({n}){k} - 2\ln(\hat{L}),
\end{align}
where $k$ is the number of free parameters in the model, $\hat{L}$ is the maximum likelihood value, and $n$ is the number of data points (in our case $n=45$, corresponding to the 45 host haloes). For Gaussian errors $2\ln(\hat{L}) = - \chi^{2}$ plus a constant, so we use the latter quantity for simplicity in this case (as the number of subhaloes is large enough that the Poisson distribution is very close to Gaussian, and only \textit{differences} in the information criteria are meaningful, so constants do not matter). We calculate $\chi^{2}$ as
\begin{equation}
\chi^{2} = \sum_{j} \frac{(N_{\rm sub}^{j, \rm meas} - N_{\rm sub}^{j, \rm pred})^{2}}{\sigma _{j}^{2}} ,
\end{equation}
where $N_{\rm sub}^{j, \rm meas}$ is the measured total subhalo abundance in the simulation above the resolution limit $V_{\rm max}^{\rm frac} = 0.065$ for the $j^{\rm th}$ halo, $N_{\rm sub}^{j, \rm pred}$ is the predicted total subhalo abundance from a given model for the $j^{\rm th}$ halo, and $\sigma _{j}$ is the uncertainty in the $j^{\rm th}$ subhalo abundance derived from the data, $\sigma _{j} = \sqrt{N_{\rm sub}^{j, \rm meas}}$; in this case we calculate $\sigma _{j}$ from the data values rather than a model both because $N$ is large (so using data values should give us a good approximation to the predicted uncertainties if we knew the true mean abundance) and to enable apples-to-apples comparisons of $\chi^2$ / AIC / BIC values between different models.

We also use the root-mean-square deviation ($\sigma_{\rm RMS}$) as a measure of the difference between predicted values from a given model (in our case $N_{\rm sub}^{\rm pred}$) and the measured values ($N_{\rm sub}^{\rm meas}$). We calculate this quantity as
\begin{equation}
\sigma_{\rm RMS} = \sqrt[]{\langle(N_{\rm sub}^{\rm meas} - N_{\rm sub}^{\rm pred})^{2}\rangle}. 
\label{equation:rmsd}
\end{equation}
In other words, $\sigma_{\rm RMS}$ is a measure of how accurately a model is able to predict the subhalo abundances measured in our simulations \citep{rms2018}. It is worth keeping in mind that the root-mean-square deviation, like chi-squared, is sensitive to outliers.

\autoref{table:AICBIC} provides the AIC, BIC, $\chi^{2}$, and $\sigma_{\rm RMS}$ values for linear versions of the models used in \autoref{sub:mw_model}, in addition to power-law version of the models (which we describe in detail below) and a model which assumes that subhalo abundance is only determined by halo mass (in which case the least-squares prediction is simply the average subhalo abundance, as halo mass is the same for all the resimulated galaxies). In this section we focus on the linear models, as those were used to select the best combinations of parameters to investigate. In our case a three-parameter model that includes linear dependence on halo spin, concentration, and shape fares better on both information criteria, $\chi^{2}$, and $\sigma_{\rm RMS}$ than the one-parameter model that incorporates concentration alone.  This three-parameter model performed better on both information criteria than almost all of the linear OLS regression models investigated, which included models linear in all the host halo parameters presented in \autoref{Section:milkyway}, quadratic terms in those quantities, and cross terms multiplying pairs of halo properties, with up to five total parameters, not including a constant term. The only exceptions were models with cross terms that become degenerate with halo properties when converted into power law form, as in the models used in \autoref{sub:mw_model}. We expect this model to be a more robust predictor for subhalo abundance. The three-parameter model without concentration is a much less robust model for predicting subhalo abundance. We include solely for comparison due to issue with current Milky Way concentration measurements as described in \autoref{Section:Adiabatic Contraction}.

\begin{table}
\centering
\begin{tabular}{llllll}
\hline
Model & AIC & BIC & $\chi^{2}$ & $\sigma_{\rm RMS}$ \\ \hline
\textbf{Linear} & & & & \\
One-para. & 146.26 & 149.88 & 142.26 & 22.34 \\ 
Three-para.&  136.26 & 143.49 & 128.26 & 21.50 \\
Three-para.\ No $c_{\rm NFW}$ &  462.11 & 469.34 & 454.11 & 39.92 \\
\hline
\textbf{Power law} & & & & \\
One-para. & 142.85 & 146.46 & 138.84 & 22.17 \\ 
Three-para. & 125.62 & 132.85 & 117.62 & 20.59\\
Three-para.\ No $c_{\rm NFW}$ & 450.16 & 457.39 & 442.16 & 39.21 \\
\hline
Mass-only model & 687.24 & 698.05 & 685.24 & 49.69 \\
\hline
\end{tabular}
\caption{Table of Akaike Information Criterion (AIC)  Bayesian Information Criterion (BIC), $\chi^{2}$, and $\sigma_{\rm RMS}$ values for the models used to estimate subhalo abundance for the Milky Way within this paper. Values are first listed for linear models which utilize one halo parameter (concentration) or three (concentration, spin parameter, and shape); values are then listed for power-law models incorporating the same quantities. We also provide statistics for a model which does not incorporate any host halo parameters, which implies that subhalo abundance is based on mass alone (as all of the resimulated haloes have roughly the same mass). The three-parameter models have lower values for AIC, BIC, $\chi^{2}$, and $\sigma_{\rm RMS}$ in all cases compared to a one-parameter model like that used in \citet{mao2015}, indicating that such models provide a better fit to the data even when penalizing them for their extra free parameters. We present results based on both one-parameter and three-parameter models in this paper in order to test the robustness of our conclusions, but our results favor the three-parameter power law model as superior to the others considered, and all halo-parameter-based models are vastly superior to the results when ignoring halo properties.  Differences in AIC/BIC of $>10$ are generally considered to provide very strong evidence of a superior fit.}
\label{table:AICBIC}
\end{table}

\begin{figure}
\centering
\includegraphics[width=\linewidth]{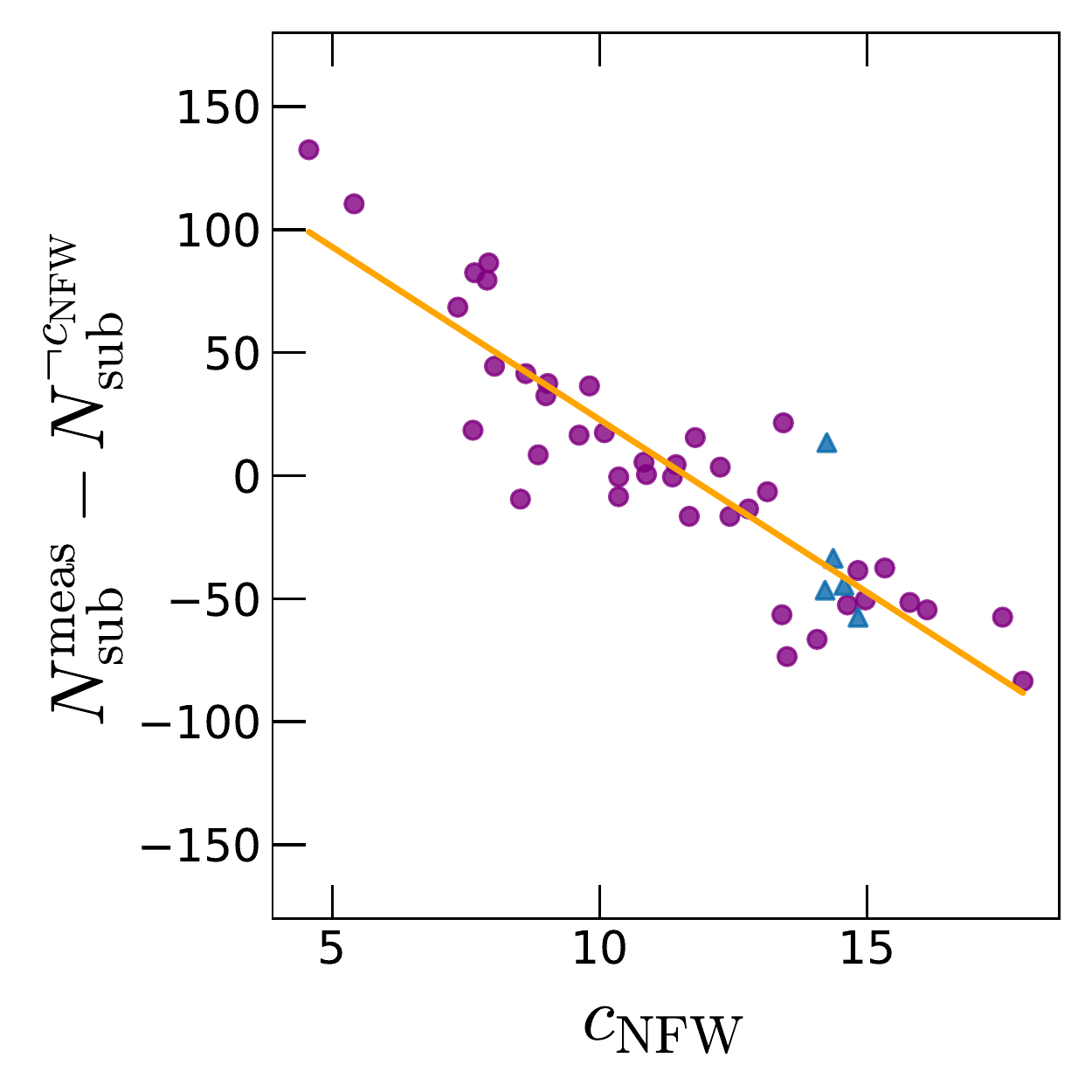}
\caption{Residual subhalo abundance as a function of the parent halo's NFW concentration parameter. In this and the figures below, the residual plotted on the y-axis is the observed $(N^{\rm sat}$ for a given halo minus the predicted $N^{\rm sat})$ for that halo from the regression model, where all terms except for the one dependent on the x-axis value are included in the prediction (in this case, this simply means that the constant term from regression has been subtracted from the total subhalo abundances above the effective resolution limit of $V_{\rm max}^{\rm frac} = 0.065$).  The over-plotted line is m$\times c_{\rm NFW}$ where m is the coefficient from the regression for the property $c_{\rm NFW}$, i.e. the term that was omitted from the regression model in calculating the residual. If the model is a good fit to the data the line should be a good representation of the plotted points.  The blue triangle points correspond to the nearest neighbors to the Milky Way in parameter space, identified as discussed in \autoref{sub:neighbors}. The tightness of the points about the line indicates that a linear function of $c_{\rm NFW}$ provides a useful prediction of subhalo number; the systematic pattern of residuals about the line at low $c_{\rm NFW}$ indicates that a non-linear dependence should give an even better fit.}
\label{fig:1param}
\end{figure}

\begin{figure*}
\centering
\includegraphics[width=\linewidth]{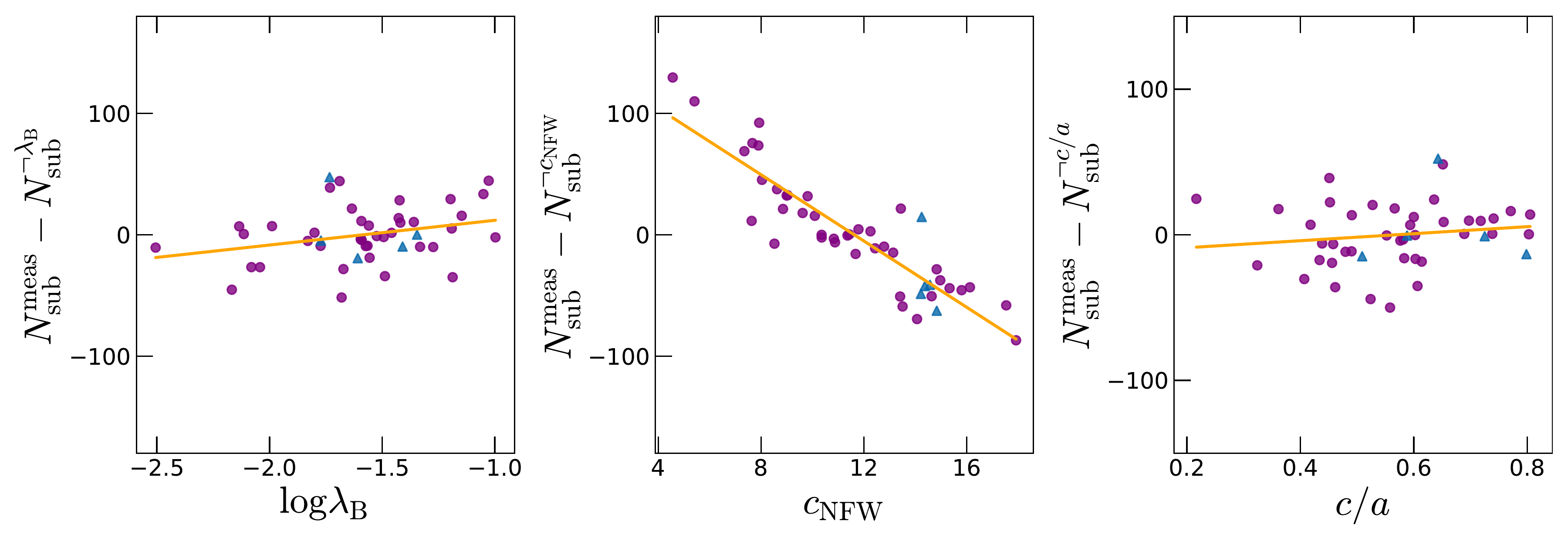}
\caption{Residual subhalo abundance as a function of the parent halo's spin, concentration, and shape. The residual plotted on the y axis is the observed $(N_{\rm sub}^{\rm meas}$ subhalo abundance for a given halo minus the predicted subhalo abundance for that halo from the regression model, where all terms except for the one dependent on the x axis value are included in the prediction ($N_{\rm sub}^{\neg x_{i}})$). The over-plotted line in each panel corresponds to the best-fit regression coefficient for a given parameter times the value of that parameter (from left to right: $\lambda_{\rm B}$, $c_{\rm NFW}$, or $ c/a$). This corresponds to the prediction of the regression model for the dependence on each quantity, with dependence on all other parameters subtracted off (as well as the constant term of the fit). The blue triangle points indicate the five nearest neighbors to the Milky Way, identified as discussed in \autoref{sub:neighbors}. The results show that a linear model based on $\lambda_{\rm B}$, $c_{\rm NFW}$, and $c/a$ provides improved predictions for subhalo abundance compared to one which depends on concentration alone.}
\label{fig:3param}
\end{figure*}

Visual depictions of the one-parameter and three-parameter OLS fits are shown in \autoref{fig:1param} and \autoref{fig:3param}. The scatter plots in each panel shows the residual value of the number of subhaloes when the prediction from a linear model is subtracted, \textit{excluding that model's dependence on the independent variable shown in that panel.} We refer to this as $N_{\rm sub}^{\rm meas} - N_{\rm sub}^{\neg x_{i}}$ where $N_{\rm sub}^{\rm meas}$ is the number of total subhaloes per host halo across the simulations above the effective resolution limit of $V_{\rm max}^{\rm frac} = 0.065$ and $N_{\rm sub}^{\neg x_{i}}$ is the predicted subhalo number, excluding the effect of host halo property $x_{i}$. In \autoref{fig:1param}, this means that what is plotted on the y axis is the number of subhaloes minus the constant term of the fit; in \autoref{fig:3param} the constant term and the dependence on all parameters but the one plotted are removed.

The over-plotted line in each panel is $ m_i\times x_i$, where $m_i$ is the coefficient from the regression fit for the property $x_i$; i.e., each line is the prediction of the regression model for the dependence on that parameter. Our methods are based off those presented in section 7 of \citet{prakash2016}. In each plot the blue triangle points correspond to the nearest neighbors to the Milky Way, as discussed in \autoref{sub:neighbors}.  These scatter plots demonstrate that the results of the regression look generally sensible, but the correlated pattern of residuals indicates that a linear fit is an imperfect representation of the data; as a result, we utilize power law dependencies in our final models. 

\begin{figure}
\centering
\includegraphics[width=\linewidth]{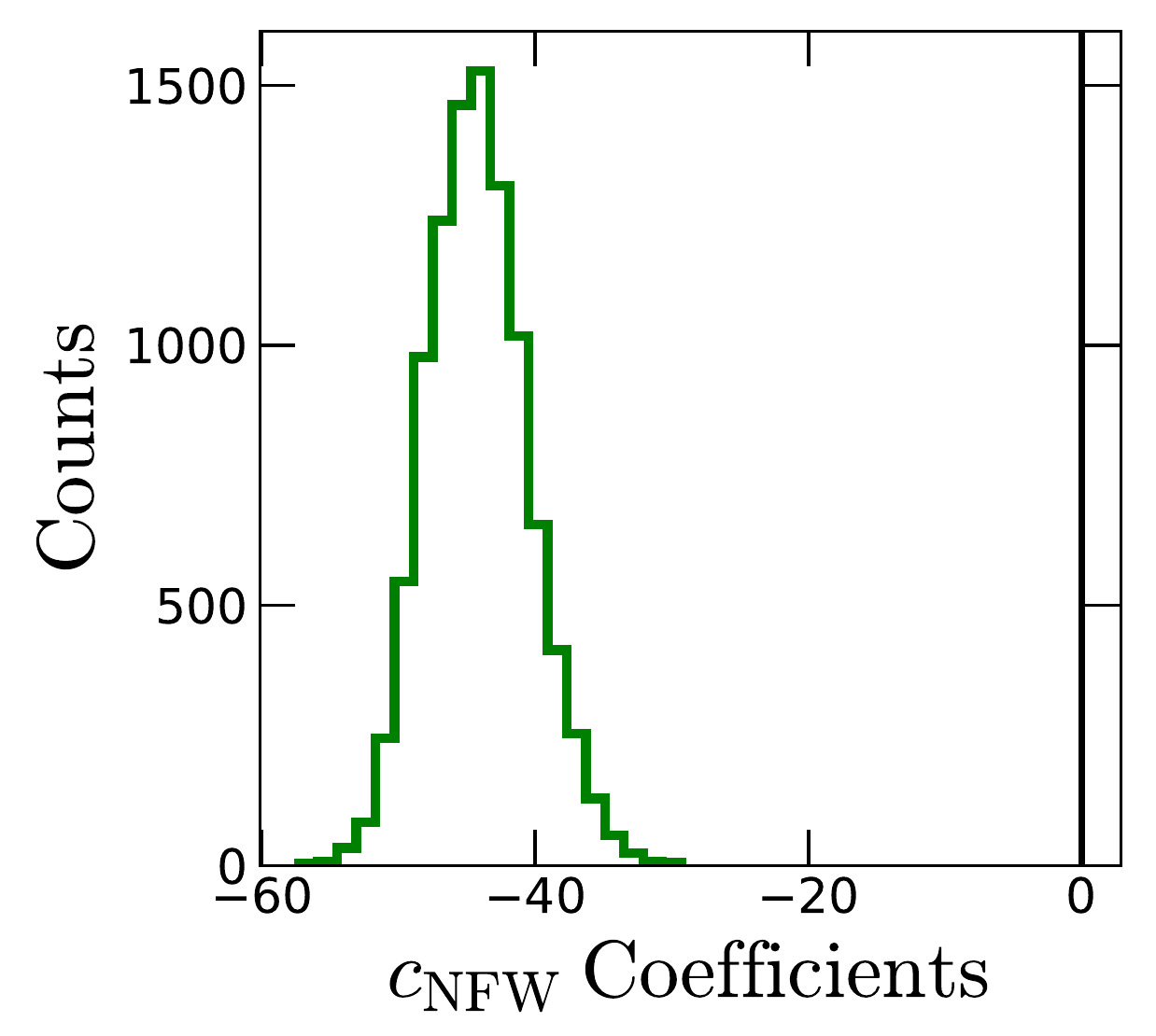}
\caption{The distribution of the coefficient of $c_{\rm NFW}$ from linear regression derived from one thousand bootstraps with replacement of the simulated halo samples. The statistics of the distribution of coefficients from bootstrapping can be used to characterize the uncertainties in the regression parameters. In this case, because the $c_{\rm NFW}$ coefficients are always non-zero,  we have firmly established that this parameter can be used to improve predictions of subhalo abundance.}
\label{fig:1_boot}
\end{figure}

\autoref{fig:1_boot} depicts the distribution of coefficients derived when OLS regression for the one-parameter is applied to 10,000 bootstrap re-samplings of the set of resimulated haloes. Bootstraps provide a reliable way to obtain errors from regression fitting even in the presence of covariances or incorrect error models. The distribution of regression coefficients amongst the bootstrap samples approximates the true PDFs for those parameters; we can use statistics of the distribution of these values as estimates for parameter uncertainties. In \autoref{fig:1_boot} the $c_{\rm NFW}$ coefficients are always non-zero, which indicates that the improvement to predictions for subhalo abundance from including this parameter is statistically significant.

\begin{figure*}
\includegraphics[width=\textwidth]{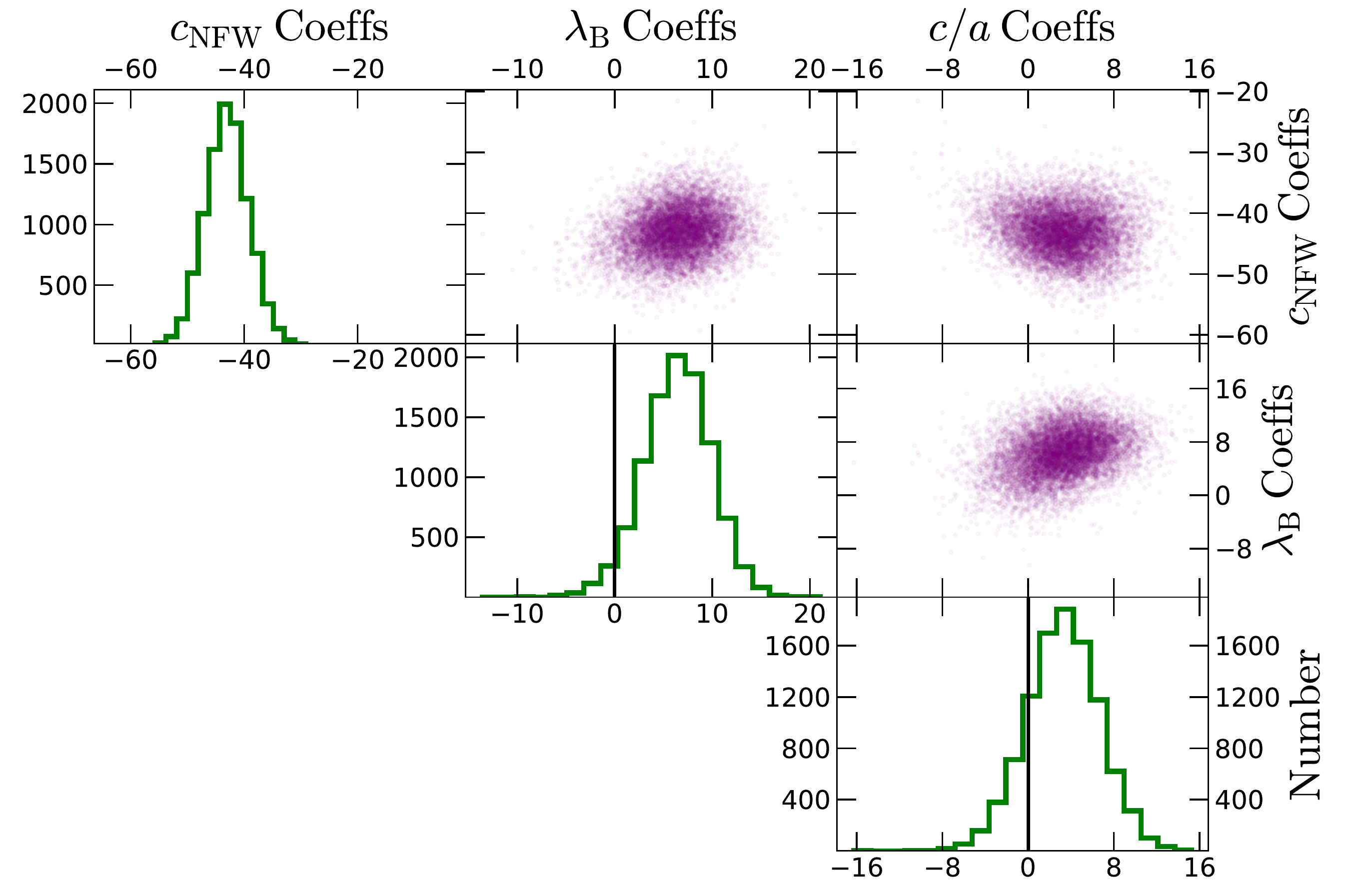}
\caption{Distributions of parameter coefficients from  bootstrap resamplings of the subhalo abundance data for the three-parameter regression depicted in \autoref{fig:3param}. The histograms correspond to how many times each value of a given coefficient occurs amongst the bootstrap samples. We can treat the distribution of coefficients from the bootstraps as a PDF for those coefficients to determine errors in regression fits.  The scatter plots show pairs of  the coefficients from each bootstrap fit plotted against each other. The scatter plot for $\lambda_{\rm B}$ vs $c/a$ shows that the coefficients have a significant correlation with each other. It is clear that not all halo parameters contribute equally to our ability to predict subhalo number; however, the AIC and BIC results provide strong evidence that a three-parameter model is superior to one that relies on concentration alone.}
\label{fig:3boots}
\end{figure*}

Similarly, \autoref{fig:3boots} shows the distribution of coefficients derived when OLS linear regression is applied to 10,000 bootstrap re-samplings with replacement of the three-parameter model. In the histograms for $\lambda_{\rm B}$, only $3.9\%$ of the coefficients are $\geq 0$ and in the histogram for $c/a$, $17.5\%$ of the coefficients are $\geq 0$. This means that these two terms are not as dominant as $c_{\rm NFW}$ in the subhalo prediction, and in particular $c/a$ is noticeably outweighed by the other two parameters (typically for significance we look for $<5\%$ of the coefficients crossing over the x = 0 line).

For multi-parameter models it is useful to also examine how the  coefficients of each parameter correlate with each other, which helps illuminate degeneracies between parameters in the fitting. As can be seen in \autoref{fig:3boots}, $c_{\rm NFW}$ and $\lambda_{\rm B}$ are the most important parameters for this model since they show little correlation, indicating that they each provide almost totally distinct information. In contrast, the scatter plot between the $\lambda_{\rm B}$ and $c/a$ coefficients amongst the bootstrap samples shows significant covariance between them. This indicates that there is likely some redundant information between the two halo parameters (which is no surprise given their relationship to each other). We conclude that although the three-parameter model provides useful predictions for subhalo abundance, not all of these parameters are contributing equally. We also point out that the more parameters there are in a model, errors on these parameters will propogate into subhalo abundance predictions. Current estimates of the Milky Way host halo properties are not very exact - therefore at current times a three-parameter model will likely have larger errors than a one-parameter model (as seen in \autoref{fig:CVF1param} and \autoref{fig:CVF3param}) but in the future when measurements are more accurate the three-parameter model is expected to provide a better prediction for subhalo abundance.

\subsection{Maximum Likelihood Fitting of a Power-Law Model for Subhalo Abundance}
\label{Section:maxlikelihood}

The linear one-parameter and three-parameter models presented above have been established to be better predictors for subhalo abundance than mass alone. In this section we investigate the improvements possible when models based upon power-law scaling relations, rather than a linear combination of parameters, are used.  We also implement improved fitting methodologies that work well even in the small-$N_{\rm sub}$ regime.  

Whereas we can write a linear model for subhalo abundance as
\begin{equation}
N_{\rm sub}^{\rm pred} = k+\sum_{i}\alpha_{i}\times x_{i},
\end{equation}
where $k$ corresponds to the intercept of a regression fit and $\alpha_{i}$ is the coefficient for the $i^{\rm th}$ halo parameter $x_i$ used in the regression (e.g., $c_{\rm NFW}$), we can similarly define a power-law model as
\begin{equation}
N_{\rm sub}^{\rm pred} = k\prod_{i}x_{i}^{\alpha_{i}}; 
\label{equation:power}
\end{equation} 
in this case the $\alpha_{i}$ are the exponents of the parameters $x_{i}$, rather than their linear coefficients.

In the previous appendices, we investigated subhalo abundance models for a case where the net number of subhaloes per halo was large, such that the Poisson distribution may be closely approximated by a Gaussian and the assumptions of least-squares regression apply. However, for higher velocity thresholds where there may be only a few subhaloes per halo, this assumption breaks down. Instead, we adopt a Poisson distribution-based maximum likelihood approach which provides secure results even in this domain. For Poisson-distributed counts the natural logarithm of the likelihood for the number of subhaloes of the $j$th halo, $\ln{\rm P(N_{\rm sub}^{\rm meas}|N_{\rm sub}^{\rm pred})}$, is given by the formula
\begin{equation}
\ln{\rm P (N_{\rm sub}^{j,\rm meas} | N_{\rm sub}^{j,\rm pred})} = N_{\rm sub}^{j,\rm meas}\ln{N_{\rm sub}^{j,\rm pred}}-N_{\rm sub}^{j,\rm pred}-\ln{\Gamma(N_{\rm sub}^{j,\rm meas}+1)},
\label{eq:poisson}
\end{equation}
where $N_{\rm sub}^{j,\rm meas}$ is the observed number of subhaloes for a given velocity threshold for the $j^{\rm th}$ host; $N_{\rm sub}^{j,\rm pred}$ is the predicted number of subhaloes for the $j^{\rm th}$ host from a given model and the same velocity threshold; and $\Gamma$ indicates the standard Gamma function.  We can maximize the likelihood of a model given the set of halo simulations by maximizing the product of their individual likelihoods (as they are independent draws from the underlying distributions the net likelihood is the product of the individual likelihoods).  However, this is equivalent to maximizing the sum of the values of the log likelihoods for each halo, i.e., $\sum_{j} \ln{\rm P(N_{\rm sub}^{\rm j,meas} | N_{\rm sub}^{\rm j,pred})}$, so we do the latter. 

Our fit values of $k$ and the power law exponents ($\alpha_{i}$) correspond to the values which maximize the total log likelihood, $\sum \ln{\rm P(N_{\rm sub}^{\rm meas}|N_{\rm sub}^{\rm pred})}$.  We determine this values by minimizing the negative of the total log likelihood using the $\tt scipy.optomize.minimize()$ function \citep{scipy,ipython} with the Nelder-Mead solver \citep{neldermead}. This function requires an initial guess which we construct by linear regression for the $\ln$ of $N_{\rm sub}^{\rm meas}$ in terms of the $\ln$'s of the $x_{i}$. The intercept of this linear fit should correspond to the $\ln$ of the $k$ parameter in \autoref{equation:power}, while the power-law exponents $\alpha_{i}$ should correspond to the linear coefficients of this regression. 

Goodness-of-fit statistics for the power-law Poisson Maximum Likelihood fits of both the one-parameter model and three-parameter model for the total subhalo abundance above the effective resolution limit of $V_{\rm max}^{\rm frac} = 0.065$ are given in \autoref{table:AICBIC}. In both cases a power-law  model provides a better fit than the equivalent linear model. The $\Delta$AIC and $\Delta$BIC for the linear versus the power-law models are $>3$ in the case of  one-parameter models and $>\sim 10$ for three-parameter models, indicating that in each case the power-law model provides a superior representation of the data.

\begin{figure*}
\includegraphics[width=\textwidth]{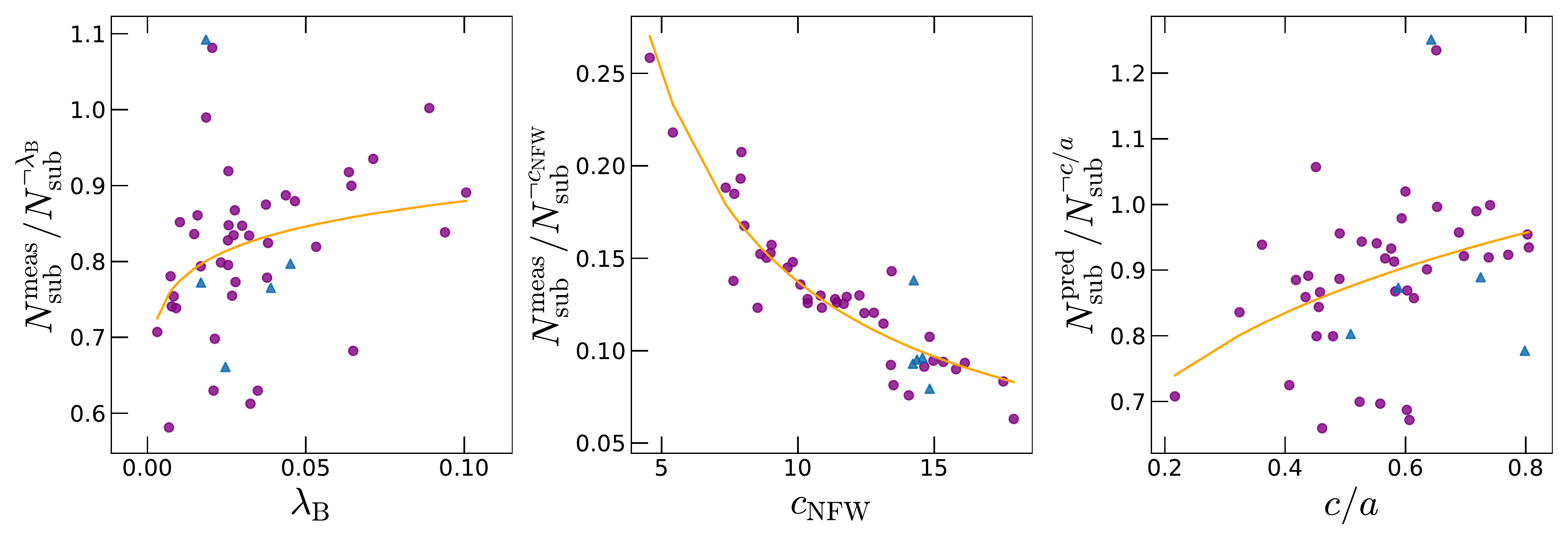}
\caption{Illustration of the three-parameter power-law model for subhalo abundance as a function of the parent halo's spin, concentration, and shape (from left to right: $\lambda_{\rm B}$, $c_{\rm NFW}$, and $ c/a$). The y-axis shows the observed $(N_{\rm sub}^{\rm meas}$ subhalo abundance for a given halo divided by the predicted subhalo abundance for that halo from a power-law model as described in \autoref{equation:power}, where all terms except for the one dependent on the x axis value are included in the prediction ($N_{\rm sub}^{\neg x_{i}})$). The over-plotted orange line in each panel corresponds to the parameter shown on the x-axis raised to the power of the best-fit exponent from our Poisson maximum likelihood fit. The blue triangle points indicate the five nearest neighbors to the Milky Way, identified as discussed in \autoref{sub:neighbors}. This result shows that a power-law is a better fit to the data than a linear fit, and that $c_{\rm NFW}$ has the greatest predictive power for subhalo abundance.}
\label{fig:3parampower}
\end{figure*}

Similarly to \autoref{fig:1param}, and \autoref{fig:3param}, we can plot the equivalent of a residual plot for the three-parameter power law model, which we provide in \autoref{fig:3parampower}. In this case the y-axis differs, as for a power-law model ratios, not differences, are more meaningful.  Hence we plot $N_{\rm sub}^{\rm meas}/N_{\rm sub}^{\neg x_{i}}$, where $N_{\rm sub}^{\neg x_{i}}$ is still the predicted subhalo abundance without including the host parameter $x_{i}$, but now using a power-law model instead of a linear one. Over-plotted in orange is the result of the power-law fit for the quantity plotted in a given panel, $x_{i}^{\alpha_i}$. For reference we also indicate the five nearest neighbors to the Milky Way as blue triangle points. Comparing \autoref{fig:3parampower} to \autoref{fig:3param}, it is evident that the power-law Poisson maximum likelihood regression does a better job than the OLS linear regression as a predictor for subhalo abundance, and once more $c_{\rm NFW}$ is the parameter that has the greatest predictive power.

\begin{figure*}
\includegraphics[width=\textwidth]{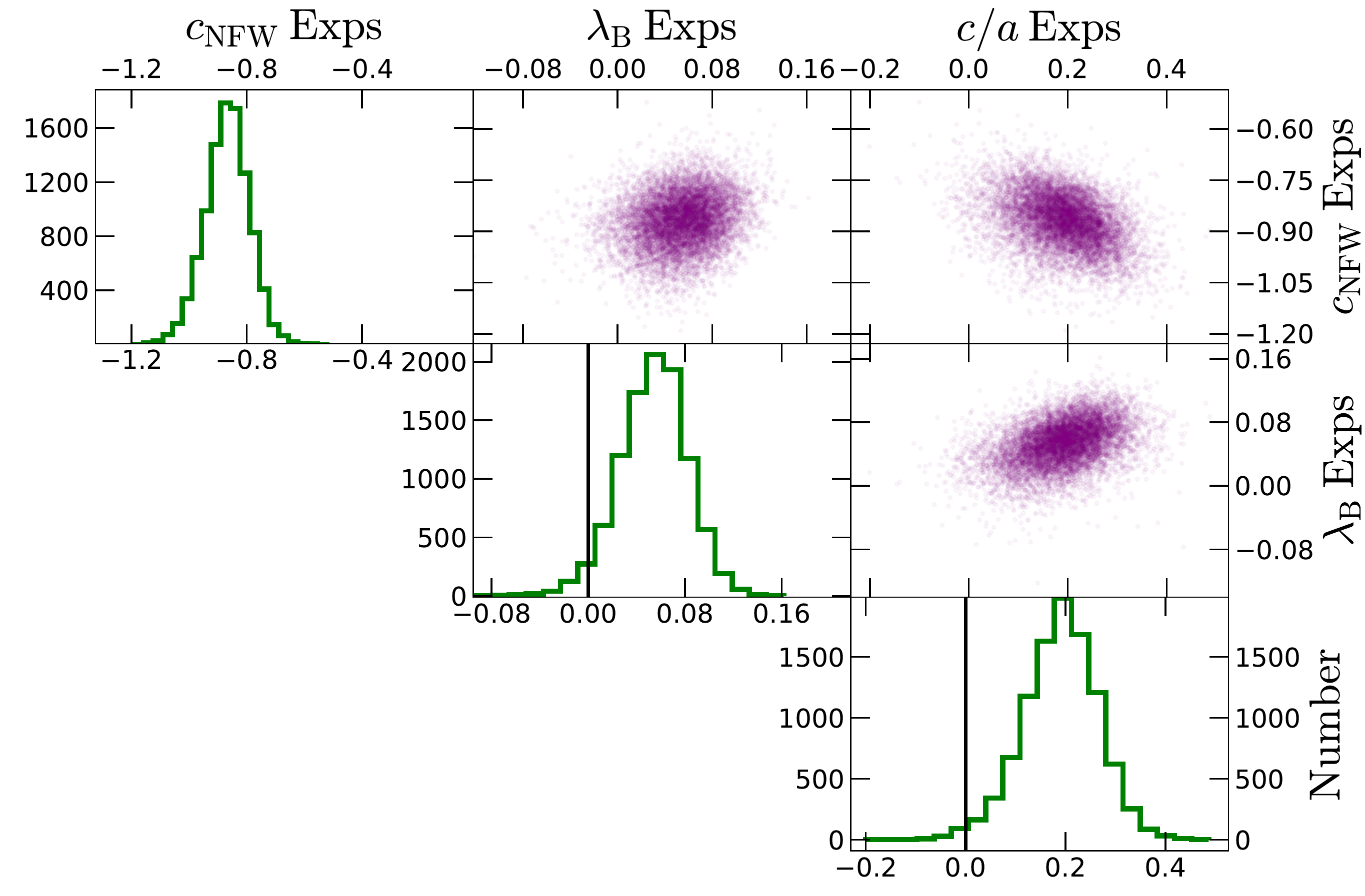}
\caption{Distributions of parameter exponents from  bootstrap resamplings of the subhalo abundance data for the three-parameter power-law Poisson maximum likelihood fit depicted in \autoref{fig:3parampower}. The histograms correspond to how many times each value of a given coefficient occurs amongst the bootstrap samples. The scatter plots show pairs of  the coefficients from each bootstrap fit plotted against each other. The scatter plot for $c/a$ vs $c_{\rm NFW}$ and $c/a$ vs $\lambda_{\rm B}$ show that the coefficients have a correlation with each other. This implies that $c/a$ is not contributing as much as the other two parameters in predicting subhalo number. However, the statistics in \autoref{table:AICBIC} show that the three-parameter model is still the superior model for predicting subhalo abundance for Milky Way-mass dark matter haloes.}
\label{fig:3bootspower}
\end{figure*}

To determine the errors in the exponents from the power-law Poisson maximum likelihood fit we again use bootstrap resampling.  Specifically, we produce 10,000 bootstrap re-samples with replacement of the ensemble of host haloes and perform the fit for each sample. \autoref{fig:3bootspower} displays the results of this process for the power-law fits; it may be compared to \autoref{fig:3boots}. 
The dependence upon halo concentration is strongest, while the dependence upon spin is relatively weak and covariant with the shape dependence.  Unlike in the case of the linear model (\autoref{fig:3boots}) there appears to be a significant degeneracy between $c_{\rm NFW}$ and $c/a$.

In order to predict the cumulative subhalo abundance for the Milky Way as a function of fractional velocity, we carry out this fitting procedure for a series of values for the minimum $V_{\rm max}^{\rm frac}$.  Specifically, our procedure is as follows:

(1) We select twenty logarithmically-spaced values of $V_{\rm max}^{\rm frac}$ to serve as the minimum fractional velocity for each of twenty bins, and determine the total number of subhaloes above every minimum value for each host halo. 

(2) For each of the twenty minimum values of $V_{\rm max}^{\rm frac}$ we fit a power-law model for predicting $N_{\rm sub}$ of the form given by \autoref{equation:power} by maximizing the total Poisson likelihood across all 45 host haloes (using \autoref{eq:poisson}), as described above.

(3) We obtain a prediction for the abundance of subhaloes for the Milky Way above each fractional velocity threshold by evaluating the corresponding power-law fit with the Milky Way's estimated halo properties (described in \autoref{Section:milkyway}) substituted in.  That is, given the $\alpha_{i}$'s and $k$ values from a Poisson maximum likelihood fit, we can predict the cumulative subhalo number for the Milky Way corresponding to each minimum value of $V_{\rm max}^{\rm sat}/V_{\rm max}^{\rm host}$ from the equation 
\begin{equation}
N^{\rm sat}(>V_{\rm max}^{\rm frac}) = k\prod_{i}x_{i,MW}^{\alpha_{i}},
\label{equation:MWnsat}
\end{equation}
where $x_{i,MW}$ is the estimated value of the $i$'th halo property for the Milky Way's host halo.  These predictions correspond to the solid purple curves in \autoref{fig:CVF1param}, \autoref{fig:CVF3param}, and \autoref{fig:CVF3param_noc}.

(4) We then perform bootstrap resampling amongst the host haloes in order to calculate the uncertainty in our power-law fits, and in the parameter estimates for the Milky Way. For each bootstrap we perturb the measured Milky Way host halo properties by a random draw from a normal distribution with standard deviation set by that property's estimated error. The case of concentration is a bit unique because our concentration estimate does not have symmetric errors. We instead build two half-gaussians with the same mean and construct the CDF of this distribution. With this we can use the PDF as a look-up table and generate random variables that correspond to that value of concentration. At the same time we select a new sample of the simulated host haloes with replacement. With this new sample we then refit for the power-law model parameters via Poisson maximum likelihood and evaluate each model with the corresponding estimated Milky Way halo properties for each bin. The resulting $68\%$ and $95\%$ confidence intervals of the cumulative velocity functions for each bootstrap sample correspond to the semitransparent purple regions in \autoref{fig:CVF1param}, \autoref{fig:CVF3param}, and \autoref{fig:CVF3param_noc}.

(5) Last, with this bootstrapped set of samples we can determine the impact of Poisson scatter on the range of values possible for the Milky Way. For each bootstrap, in each bin of $V_{\rm max}^{\rm frac}$ we compute the predicted number of subhaloes for the Milky Way per bin for a non-cumulative satellite abundance. Then we draw randomly from a Poisson distribution using this non-cumulative predicted abundance as the mean. Last, these Poisson values are cumulatively summed in order to determine the error corresponding to the scatter of the 45 haloes about the model prediction. The $68$ and $95$ percent confidence regions for this result are depicted in orange in \autoref{fig:CVF1param}, \autoref{fig:CVF3param}, and \autoref{fig:CVF3param_noc}. Applying a log-normal error distribution yields very similar results to the Poisson scatter utilized here. We also compare the errors for the total cumulative number of subhaloes from the bootstraps alone and the additional impact of Poisson scatter in \autoref{fig:histerror}. The incorporation of Poisson noise leads to a much wider spread in predicted subhalo abundance at higher $V_{\rm max}$ and a comparable spread at low $V_{\rm max}$. Percentile confidence regions for these two error models on total subhalo abundance normalized by the mean measured subhalo abundance are shown numerically in \autoref{table:errors} and visually in \autoref{fig:histerror}.

The end result of this process is both a best-fit, best-estimate cumulative velocity function for the Milky Way and a set of additional CVFs whose distribution reflects the uncertainties in fitting power-law models, uncertainties in the parameters of the Milky Way host halo, and uncertainties due to Poisson scatter about the relation. 

\subsection{Fitting Functions for the Subhalo Cumulative Velocity Function}
\label{sub:generalfunc}
Our best-fit models for the cumulative velocity function of subhaloes as a function of host halo parameters can be approximately described by a simple set of formulae. Our goal is to provide easily-computed values which can be substituted into \autoref{equation:MWnsat} to predict the cumulative velocity function for any dark matter halo. We take advantage of the fact that (due to the self-similarity of dark matter haloes of different masses and the virial scaling relations), the number of subhaloes above a given $V_{\rm max}^{\rm frac}$ should have minimal dependence on halo mass.  That is, we expect the number of satellites above a given \textit{fraction} of a halo's $V_{\rm max}$ to be similar regardless of halo mass, but in a more massive halo, the satellites that are above that fractional threshold will also be more massive (this also leads to the conclusion that there will be more total satellites above a fixed minimum $V_{\rm max}$ in more massive haloes, roughly proportional to the halo mass).

To define an approximate model for the CVF, we need to specify values of $k$ and $\alpha_i$ as a function of the minimum subhalo fractional velocity considered, $V_{\rm max}^{\rm frac}$.  Specifically, the results from Poisson maximum likelihood one-parameter power-law fits can be accurately represented using a function that is linear at low velocities and quadratic at high velocities.  We have fit for this function using the Numpy $\tt curve\_fit()$ routine applied to the values of $\alpha_{c_{\rm NFW}}$ as a function of $V_{\rm max}^{\rm frac}$ obtained from the Poisson maximum likelihood results, using weights calculated from the standard deviation of the bootstrap results for this parameter at a given velocity. From this we obtain a fit:
\begin{equation}
\alpha_{c_{\rm NFW}} = \left\{\begin{array}{ll}
      0.12V^{\dagger} - 0.78, & V^{\dagger} < 0, \\
       -66.26V^{\dagger 2} - 1.02V^{\dagger} - 0.78, & V^{\dagger} \geq 0,
\end{array}
\right. 
\label{equation:cfit1}\end{equation}
where $V^{\dagger} = V_{\rm max}^{\rm frac}-0.12$. This fit provides a good representation of the dependence of the power-law exponent on velocity over the range $0.05 \leq V_{\rm max}^{\rm frac} \leq 0.25$.

 We find that a single quadratic function of $V_{\rm max}^{\rm frac}$ is sufficient for characterizing the dependence of the power-law prefactor $k$ on velocity. In order to ensure a self-consistent set of fitting functions, we must obtain new values of $k$ for each threshold velocity while forcing $\alpha_{c_{\rm NFW}}$ to have the value predicted by \autoref{equation:cfit1} for $\alpha_{c_{\rm NFW}}$. Specifically, we take $k = \langle N_{\rm sub}^{\rm meas}/N_{k=1}^{\rm sat}(>V_{\rm max}^{\rm sat})\rangle$, where $N_{\rm sub}^{\rm meas}/N_{k=1}^{\rm sat}(>V_{\rm max}^{\rm sat})$ is the satellite abundance estimate for each host halo taking k=1 and the value of $\alpha_{c_{\rm NFW}}$ from \autoref{equation:cfit1}; i.e., $N_{k=1}^{\rm sat}(>V_{\rm max}^{\rm sat}) = c_{\rm NFW}^{\alpha_{c_{\rm NFW}}}$. By a least-squares fit to the values of the logarithm of $k$ as a function of threshold velocity, we obtain
\begin{equation}
\ln{k} = 205.58(V_{\rm max}^{\rm frac})^{2}-73.66V_{\rm max}^{\rm frac}+10.93.
\label{equation:kfit1}
\end{equation}
The results of \autoref{equation:cfit1} and \autoref{equation:kfit1} can then be substituted into \autoref{equation:power} to obtain a prediction for the subhalo abundance at a given threshold in velocity based on a dark matter halo's  concentration. The results of these fitting formulae match the mean predictions for the Milky Way (corresponding to the purple line in \autoref{fig:CVF1param}) to better than $2.7\%$ RMS over the range in velocities $0.05 \leq V_{\rm max}^{\rm frac} \leq 0.25$.

We similarly have derived fitting functions which can be used to approximate our three-parameter model fits. We begin by noting that the exponents of the spin and shape parameters from our bootstrap samples are consistent with being constant at all velocities.  As a result, for our fitting formula we treat them as fixed at their median values across all bootstrap samples from the original maximum likelihood calculation, corresponding to 
\begin{align}
\alpha_{\lambda_{\rm B}} &= 0.03; \\
\alpha_{c/a} &= 0.26.
\end{align}
We then perform new fits for a new $k$ and $\alpha_{c_{\rm NFW}}$ via Poisson maximum likelihood, including spin and shape in the model but forcing their exponents to have the fixed values described above. 

Then we proceed the same as for the one-parameter model. We find that the concentration exponent for the three-parameter model can be approximated well by
\begin{equation}
\alpha_{c_{\rm NFW}} = \left\{\begin{array}{ll}
      0.10V^{\dagger} - 0.91 , & V^{\dagger} < 0, \\
       -71.86V^{\dagger 2} + 2.02V^{\dagger} - 0.91, & V^{\dagger} \geq 0,\\
\end{array} 
\right. 
\label{equation:cfit3}\end{equation}
where $V^{\dagger} = V_{\rm max}^{\rm frac}-0.11$ in this case. Once again, we use the fitting formula predictions for $\alpha^{c_{\rm NFW}}$ to estimate the $k$ term at each velocity, and then fit for $\ln{k}$ as a function of $V_{\rm max}^{\rm frac}$ via least-squares.  We then find 
\begin{equation}
\ln{k} = 208.21(V_{\rm max}^{\rm frac})^{2}-74.11 V_{\rm max}^{\rm frac}+11.47.
\label{equation:kfit3}
\end{equation}
The results from \autoref{equation:cfit3} and \autoref{equation:kfit3} can be substituted into \autoref{equation:power} to obtain a prediction based on a dark matter halo's concentration, spin, and shape for subhalo abundance above a given velocity threshold. The results of these sets of fitting formulae match the mean Milky Way prediction (the solid purple line in \autoref{fig:CVF3param}) to $6.9\%$ difference RMS for $0.05 \leq V_{\rm max}^{\rm frac} \leq 0.25$.

We note that if realizations of a cumulative velocity function incorporating Poisson statistics are desired, care must be taken to ensure that covariances between values at different velocities resulting from the use of a cumulative quantity are properly accounted for. Step five in \autoref{Section:maxlikelihood} describes the procedure which we have employed for this purpose. 

\label{lastpage}

\end{document}